\documentclass{article}
\usepackage[utf8]{inputenc}

\usepackage{amsfonts,amsmath,amssymb} 
\usepackage{booktabs} 
\usepackage{color} 
\usepackage{fancyhdr} 
\usepackage{graphicx} 
\usepackage{units} 
\usepackage{xspace} 
\usepackage[version=4, arrows=pgf-filled]{mhchem} 
\usepackage{multirow} 
\usepackage{tabularx} 
\usepackage{placeins}
\usepackage{longtable}
\usepackage{algorithm} 
\usepackage{authblk}
\usepackage{overpic}

\newcolumntype{L}[1]{>{\raggedright\arraybackslash}p{#1}}
\newcolumntype{C}[1]{>{\centering\arraybackslash}p{#1}} 
\newcolumntype{R}[1]{>{\raggedleft\arraybackslash}p{#1}}

\usepackage{array} 
\usepackage{subcaption} 
\usepackage{rotating} 
\usepackage{todonotes} 
\usepackage{caption}
\usepackage{subcaption}
\usepackage{lscape}
\usepackage{pdflscape}
\usepackage[bookmarks=true,bookmarksopen=true,bookmarksopenlevel=2]{hyperref} 
\hypersetup{
  pdfpagemode=UseOutlines,
  pdfstartview=Fit,
  pdftitle={},
  pdfsubject={},
  pdfauthor={},
  pdfkeywords={},
  pdfcreator={pdflatex}
}
\usepackage{cleveref} 
\usepackage{doi}

\numberwithin{equation}{section}
\numberwithin{figure}{section}
\numberwithin{table}{section}

\newcolumntype{V}[1]{>{\raggedright\hspace{0pt}}p{#1}} 
\newcommand{\dumux}{DuMu$^\textrm{x}$\xspace}

\newtheorem{remark}{Remark}
\title{Impact of saturation on evaporation-driven density instabilities in porous media: mathematical and numerical analysis}
\author[1]{C. Bringedal}
\author[2]{S. Kiemle}
\author[3]{C. J. van Duijn}
\author[2]{R. Helmig}

\affil[1]{Department of Computer Science, Electrical Engineering and Mathematical Sciences, Western Norway University of Applied Science, Bergen, Norway}
\affil[2]{Institute for Modelling Hydraulic and Environmental Systems, University of Stuttgart, Stuttgart, Germany}
\affil[3]{Department of Mechanical Engineering, Eindhoven University of Technology, Eindhoven, The Netherlands}

\begin{document}
\maketitle

\begin{abstract}
    Evaporation from a porous medium partially saturated with saline water, causes the salinity (salt concentration) to increase near the top of the porous medium as water leaves while salt stays behind. As the density of the water increases with increased salt concentration, the evaporation leads to a gravitational unstable setting, where density instabilities can form. Whether density instabilities form, depends on a large range of parameters like the evaporation rate and intrinsic permeability of the porous medium, but also on the water saturation. As water saturation decreases, the storage, convection and diffusion of salt also decrease, which all influence the onset of instabilities. By performing a linear stability analysis on the governing equations, we give criteria for onset of instabilities. Numerical simulations give information about the further development of these instabilities. With this knowledge we can predict whether and when density instabilities occur, and how they will influence the further development of salt concentration in the porous medium.
\end{abstract}

\setcounter{page}{1}  

\pagestyle{fancy}
\renewcommand{\sectionmark}[1]{\markright{\thesection\ #1}}
\fancyfoot{}
\fancyhead{}
\fancyhead[LE,RO]{\small\bfseries \thepage}
\fancyhead[LO]{\small\bfseries \rightmark}
\renewcommand{\headrulewidth}{0.5pt} 

\section{Introduction}

Soil salinization is a major environmental risk as it hampers plant growth and affects biological activity \cite{salinityreview2016}. As water evaporates from the soil, the dissolved salts stay behind, causing the salinity of the saturating water to gradually increase. This increased salinity can have a negative impact on root water uptake \cite{chaves2008plant}. If the salinity increases such that the solubility limit of the salt is reached, the salt will precipitate and potentially form a salt crust that disconnects the soil from the atmosphere \cite{chen1992evaporation,jambhekar,emna2017,piortrowski2020crust}. This salt crust has a strong negative impact on the growth conditions for many agricultural plants \cite{singh2015,pitman2020salinity}.

Evaporation of water hence causes an increase in salt concentration in the remaining water that is (partially) saturating the soil \cite{allison1985}. As the evaporation mainly takes place near the top of the soil, the subsequent increase in salt concentration is therefore stronger here. Although diffusion will cause concentration differences to smoothen out, there will be an interim phase where the concentration is larger at the top of the porous medium. As the concentration of the dissolved salts increase, the density of the solvent (the water) will also increase \cite{GengInfluence}. This creates a gravitational unstable setting with heavier, denser water lying on top of lighter, less dense water \cite{gilman1996influence}. As the soil is permeable, instabilities in the form of downwards flow caused by these density differences can form \cite{WoodingSalt1,WoodingSalt2}. However, whether such density instabilities occur, depends on whether the density difference is large enough to overcome the resistance of the porous medium and therefore trigger downwards flow. If the density instabilities form, the downwards convective transport of salt can give a reduction of the salt concentration near the top of the porous medium.

The occurrence of density instabilities is shown by the method of linear stability. This method has been applied to a wide range of problems where a density difference causes a gravitationally unstable setting \cite{nieldbejan2017}. By prescribing a fixed salt concentration at the top of the porous medium, the onset of instabilities was analyzed in \cite{Elenius2012,lasser2021stability,Riaz2006,vanduijn2019stability}. These studies found that density instabilities in the form of fingers form when the density difference is strong enough to overcome the resistance of the porous medium. This is quantified through a Rayleigh number, where a Rayleigh number larger than a critical threshold means that density instabilities can occur. 

When considering evaporation from a porous medium, the salt concentration near the top of the porous domain will gradually increase with time as the water evaporates and salts stay behind. This corresponds to a Robin-type of boundary condition for the salt concentration, while the above-mentioned studies considered a Dirichlet-type of boundary condition. Density instabilities with Robin-type of boundary conditions have been investigated using a linear stability analysis earlier, see e.g.~\cite{barletta2009robin,hattori2015robin}. 
In our previous work, we considered evaporation from a fully saturated porous medium \cite{BringedalSatInstab,KlokerApproaches}. As the porous medium was assumed to remain fully saturated, we could consider single-phase flow, and a standard convection-diffusion equation for the dissolved salt concentration. By performing a linear stability analysis we found criteria for onset of instabilities and corresponding onset times, expressing when the density difference is large enough to overcome the resistance of the porous medium. Onset times were therefore much later when the medium's permeability was lower \cite{BringedalSatInstab}. Comparison with numerical simulations of the full system of equations confirmed this behavior and gave comparable onset times when a similar type of perturbation was used \cite{BringedalSatInstab,KlokerApproaches}.

The assumption that the porous medium remains fully saturated during evaporation, was limiting the applicability of the previous studies \cite{BringedalSatInstab,KlokerApproaches}. In this work, we therefore extend the analysis to partially saturated porous media by applying Richards' equation. Then, the water saturation of the porous medium is a variable, which can change due to the evaporation at the top of the domain and due to the water flow within the porous medium. The convection-diffusion equation for the salt concentration is accordingly adjusted and will be influenced by the changing saturation through the storage term, the convection term and the diffusion term \cite{Helmig2011}. All these can potentially have a strong influence on the onset of density instabilities. There is hence a strong coupling and interplay between the varying saturation and the onset of instabilities.

Gravitational (in)stability of special solutions (steady state, travelling waves) was addressed in \cite{egorov2003stability,ursino2000linear,van2004steady}. These papers consider Richards' equation in the unsaturated setting, but with constant fluid density. Then, Richards' equation is found to be unconditionally stable in the sense that saturation fingers cannot develop in the standard formulation \cite{egorov2003stability}. For a discussion about instabilities in multiphase porous-media flow, we refer the reader to the review by DiCarlo \cite{dicarlo2013stability}. The models considered in these papers differ fundamentally from the model we propose in the current study. Here we include the influence of varying density to get an interplay between density and saturation changes during evaporation.

In this work we therefore study the onset and development of density instabilities caused by evaporation from a partially saturated porous medium. The goal of this paper is to find onset criteria and onset times of density induced gravitational instabilities as a function of the system parameters, and address the further development of the instabilities. The method of linear stability analysis is used to give estimates for the onset of instabilities and for the time of their appearance. This analysis relies on some restrictive assumptions to be performed on this highly coupled model, but has the advantage that it can give criteria for a large range of parameters at low computational costs. The linear stability analysis can furthermore be used to investigate the influence of the saturation by addressing changes in storage, convection and diffusion independently. We use numerical simulations to find onset times and to address the development of the instabilities and in particular the development of the salt concentration near the top of the porous medium. 

This paper is organized as follows. In Section \ref{sec:model} we present the model equations for fluid flow and salt concentration in a partially saturated porous medium, along with initial and boundary conditions used in this study. Section \ref{sec:lsa} presents the linear stability analysis, while Section \ref{sec:numModel} presents the numerical solver strategy. The results from the linear stability analysis and of the numerical simulations are presented in Section \ref{sec:results}, before conclusion and outlook are given in Section \ref{sec:final}.

\section{Mathematical Model}\label{sec:model}
\begin{figure}[ht!]
	\centering
	\includegraphics[width=0.75\textwidth]{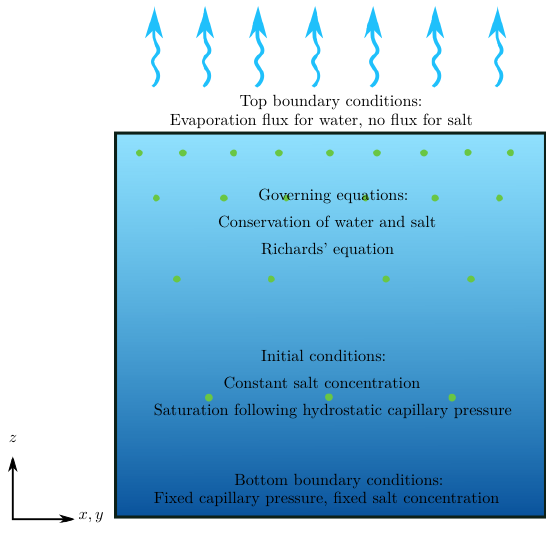}
	\caption{Sketch of domain with processes governing evaporation from a partially saturated porous medium}
	\label{fig:model}
\end{figure}

The following section presents the underlying physical assumptions and boundary conditions that outline the mathematical model to describe evaporation from a partially saline-water-saturated porous medium. The domain together with most important model choices are given in Figure \ref{fig:model}.

Since the medium is partially saturated, we need to account for two phases: gas (air) and liquid (water). Salt is assumed to be dissolved in the liquid phase and we employ the Richards' equation to describe the flow of the liquid phase. Although thermal effects are generally important when considering evaporation, we consider an isothermal model as our main interest are the changes in the concentration of dissolved salt and subsequent density instabilities.
As we are interested in time frames before the dissolved salt would potentially precipitate and become part of the solid matrix, we consider the porosity and permeability to be constant with time. As we are interested in density-driven gravitational flow, we employ the Boussinesq approximation. We consider a domain $\Omega$, which is vertically bounded between $z=0$ and $z=H$. Horizontally, the domain will either be unbounded or has vertical sidewalls at $x,y=0$ and $x,y=\ell$. The governing equations inside $\Omega$, in terms of (water) saturation $S_\mathrm{w}\ [\text{m}^3\ \text{m}^{-3}]$ and salt mole fraction $\mathsf{x}_\mathrm{w}^\mathrm{NaCl}\ [\text{mol}\ \text{mol}^{-1}]$ as unknowns, are:

\begin{align}
	\phi\partial_tS_\mathrm{w} + \nabla\cdot\mathbf{Q_\mathrm{w}} &= 0,\label{eq:mass}\\
	\phi\partial_t(S_\mathrm{w} \mathsf{x}_\mathrm{w}^\mathrm{NaCl}) + \nabla\cdot(\mathbf{Q}_\mathrm{w} \mathsf{x}_\mathrm{w}^\mathrm{NaCl}-\phi^{3/2} f(S) D^\mathrm{NaCl}_\mathrm{w}\nabla \mathsf{x}_\mathrm{w}^\mathrm{NaCl}) &=0, \label{eq:NaCl} \\
	\mathbf{Q}_\mathrm{w} = \frac{ k_\mathrm{rw}(S_\mathrm{w})}{\mu_\mathrm{w}(\mathsf{x}_\mathrm{w}^\mathrm{NaCl})}K(\nabla p_c - \rho_\mathrm{w}(\mathsf{x}_\mathrm{w}^\mathrm{NaCl}) g\mathbf{e}_z). \label{eq:Darcy}&
\end{align}

The first two equations express the mass conservation of the water phase and of the salt concentration, while the third equation is the extended Darcy's law for multiphase flow \cite{Helmig2011}. In the above equations, subscript $\mathrm{w}$ indicates the water phase, while superscript $\mathrm{NaCl}$ indicates the salt. Furthermore, $\phi\ [\text{m}^3\ \text{m}^{-3}]$ is porosity, and $\mathbf Q$ $[\text{m}\ \text{s}^{-1}]$ is the Darcy velocity. Note that the use of mole fraction for the salt concentration would hint towards that the water mole fraction in the water phase would be slightly less than 1, and that the two mole fractions sum up to 1. As we are interested in low salt concentrations (below the precipitation limit), we do not consider the water mole fraction as a variable. The diffusive term in \eqref{eq:NaCl} employs an effective diffusion law modified after \cite{Millington1961}, where $f(S_\mathrm{w})=S_\mathrm{w}^{7/2}$. The coefficient $D_\mathrm{w}^\mathrm{NaCl}$ is the intrinsic diffusion of component NaCl in the water phase. The coefficient $K\ [\text{m}^2]$ is the intrinsic permeability of the porous medium, and $g\ [\text{m}\ \text{s}^{-2}]$ is the gravity acceleration. Finally, $k_\mathrm{rw}(S_\mathrm{w})\ [-]$ is the relative permeability, $p_c(S_\mathrm{w})\ [\text{Pa}]$ the capillary pressure, while $\mu_\mathrm{w}(\mathsf{x}_\mathrm{w}^\mathrm{NaCl})\ [\text{Pa}\ \text{s}],\rho_\mathrm{w}(\mathsf{x}_\mathrm{w}^\mathrm{NaCl})\ [\text{kg}\ \text{m}^{-3}]$ are the viscosity and density of the water phase as given functions of the salt concentration. Note that $\mathbf{e}_z$ denotes the unit vector pointing upwards.

Although we let the water density vary with salt concentration, we have in \eqref{eq:mass} and \eqref{eq:NaCl} used the Boussinesq approximation, which allows us to neglect density differences unless they occur in the gravity term of Darcy's law \eqref{eq:Darcy}. We also suppose that the salt mole fraction stays below the solubility limit. 
Hence, these model equations are suitable to address onset of instabilities before salt precipitates.
As we are interested in density-driven flow, we can this way allow for gravitationally unstable flow via \eqref{eq:Darcy}, while at the same time employing simpler mass conservation equations \eqref{eq:mass} and \eqref{eq:NaCl}.

Since we have partially saturated flow, we need closure relationships for the capillary pressure and the relative permeability. Here we use the van Genuchten expressions \cite{vanGenuchten1980} 

\begin{align}
	p_\mathrm{c}(S_\mathrm{we}) &= \frac{1}{\alpha}\left( S_\mathrm{we}^{-1/m}-1 \right)^{1/n}
	\mathrm{with}~ S_\mathrm{we} =\frac{S_\mathrm{w} - S_\mathrm{wr}}{1-S_\mathrm{wr}} \text{ and }
	m=1-\frac{1}{n},
	\label{eq:pcSw} \\
	k_\mathrm{rw}(S_\mathrm{we}) & = S_\mathrm{we}^{L} \left(1-(1-S_\mathrm{we}^{1/m})^m\right)^2,
	\label{eq:kr_sw_realtionship}
\end{align}

where $\alpha\ [\text{Pa}^{-1}], m,n, L$ are constants depending on the type of porous medium. Furthermore, $S_\mathrm{wr}$ is the residual water saturation, which is also a constant depending on the porous medium.

The water density and viscosity are assumed to vary with the salt concentration using the empirical relations formulated in \cite{Batzle1992}. These relations also include a dependence on water pressure and temperature. This we disregard by fixing a constant water pressure of $10^5 \text{ Pa}$ and a temperature of $20^\circ \text{ C}$ to focus on the dominating influence of varying salt concentration only:

\begin{align}
	\label{eq:densityDumux}
	\rho_\mathrm{w}(X^\mathrm{NaCl}_\mathrm{w}) =&\ 998.203 + 1000  X^\mathrm{NaCl}_\mathrm{w}(0.670804 + 0.373854X^\mathrm{NaCl}_\mathrm{w})\\
	\mu_\mathrm{w}(X^\mathrm{NaCl}_\mathrm{w}) =&\ \frac{1}{1000}\Big(0.1+0.333X^\mathrm{NaCl}_\mathrm{w} \nonumber \\ &+ (1.65+91.9(X^\mathrm{NaCl}_\mathrm{w})^3)\cdot \mathrm{exp}\big( -4.614( (X^\mathrm{NaCl}_\mathrm{w})^{0.8}-0.17)^2-0.494\big) \Big) \label{eq:variableviscosity}
\end{align}

where $X^\mathrm{NaCl}_\mathrm{w}\ [\text{kg}\ \text{kg}^{-1}]$ is the salt \emph{mass} fraction.

At the bottom boundary, we assume that the porous medium is connected to a water reservoir at a fixed (capillary) pressure $p_B$, which contains a salt concentration that is equal to the constant initial concentration $\mathsf{x}_0^\mathrm{NaCl}$. At the top boundary, we assume that there is an evaporation flux $E\ [\text{m}\ \text{s}^{-1}]$ of water, but a no flux condition for salt. Therefore,

\begin{align}
	\mathbf{Q}_w|_{z=H} &= E\mathbf{e}_z, \\
	(\mathbf{Q}_\mathrm{w} \mathsf{x}_\mathrm{w}^\mathrm{NaCl}-\phi^{3/2} f(S_\mathrm{w}) D_\mathrm{w}^\mathrm{NaCl}\nabla \mathsf{x}_\mathrm{w}^\mathrm{NaCl})|_{z=H}\cdot\mathbf{e}_z &= 0,\\
	p_c|_{z=0} &= p_B,\\
	\mathsf{x}_\mathrm{w}^\mathrm{NaCl}|_{z=0} &= \mathsf{x}_0^\mathrm{NaCl}.
\end{align}

Initially, the porous medium is assumed to be partially saturated according to hydrostatic pressure following \eqref{eq:pcSw}, and a constant salt concentration. That is,

\begin{align}
	\mathsf{x}_\mathrm{w}^\mathrm{NaCl}|_{t=0} = \mathsf{x}_0^\mathrm{NaCl},\\
	p_c|_{t=0}= p_0 \text{ where } \frac{dp_0}{dz}=\rho_{\mathrm{w}}(\mathsf{x}_0^\mathrm{NaCl})g \text{ and } p_0|_{z=0} = p_B.
\end{align}

If there are sidewalls, we will apply no flow for the water phase, and no flux for the salt concentration. Hence,

\begin{align}
	\mathbf{Q}_\mathrm{w}|_{x,y=0,\ell}\cdot\mathbf e_{x,y} &= 0,\\
	\partial_{x,y} \mathsf{x}_\mathrm{w}^\mathrm{NaCl}|_{x,y=0,\ell} &=0,
\end{align}

where $\mathbf{e}_{x,y}$ denotes the unit vectors pointing in either $x$ or $y$ direction, and $\partial_{x,y}$ is first-order partial derivative with respect to $x$ or $y$.

\section{Linear Stability Analysis}\label{sec:lsa}

The model equations presented in Section \ref{sec:model} have a unique solution, which is called the ground state. 
To address the question of stability of the ground state, we perform a linear stability analysis. 
The steps of the linear stability analysis are standard. 
However, given the coupled nature of the model equations, the resulting eigenvalue problem is complex and non-standard and will need to be further simplified.

The model equations from Section \ref{sec:model} are non-dimensionalized to identify dimensionless numbers that characterize the overall behavior of the model. The ground state is perturbed using a quasi-static approach \cite{Riaz2006}, in which we assume that the ground state is changing slowly compared to the growth of instabilities.
From this we derive perturbation equations that are then linearized. The linearized perturbation equations are then finally formulated as an eigenvalue problem.
In this eigenvalue problem, time appears as a parameter through the ground state. By solving it we obtain information about the stability of the ground state, which can be translated to onset times of instabilities.

Note that the analysis is performed in a general manner and could be applied to any equation of state for the density and viscosity and also for other expressions for the relative permeability, capillary pressure saturation relationship and effective diffusion. We however require that these relations are smooth, that the capillary pressure relationship is invertible, and that the relative permeability is positive. The analysis is valid for other types of salts than NaCl and other soil types.
However, as mentioned in Section \ref{sec:model}, the analysis is only valid as long as the salt mole fraction is below the solubility limit.

\subsection{Reformulated model equations}\label{sec:lsa_eqs}
We consider slightly different model equations from Section \ref{sec:model}, by using capillary pressure $p_c$ instead of saturation $S_\mathrm{w}$ as unknown. This is done to later employ the Kirchhoff potential. The capillary pressure saturation relationship \eqref{eq:pcSw} is inverted such that 

\begin{equation}\label{eq:pcSinv}
	S_\mathrm{w} = S_\mathrm{w}(p_c) = S_\mathrm{wr}+(1-S_\mathrm{wr})(1+(\alpha p_c)^n)^{-m}.
\end{equation}

Therefore, the model equations considered in the linear stability analysis are

\begin{align}
	\phi\partial_t(S_\mathrm{w}(p_c)) + \nabla\cdot\mathbf{Q_\mathrm{w}} &= 0,\\
	\phi\partial_t(S_\mathrm{w}(p_c) \mathsf{x}_\mathrm{w}^\mathrm{NaCl}) + \nabla\cdot(\mathbf{Q}_\mathrm{w} \mathsf{x}_\mathrm{w}^\mathrm{NaCl}-\phi^{3/2}\tilde f(p_c) D^\mathrm{NaCl}_\mathrm{w}\nabla \mathsf{x}_\mathrm{w}^\mathrm{NaCl}) &=0, \\
	\mathbf{Q}_\mathrm{w} = \frac{\tilde k(p_c)}{\mu(\mathsf{x}_\mathrm{w}^\mathrm{NaCl})}K(\nabla p_c - \rho_\mathrm{w}(\mathsf{x}_\mathrm{w}^\mathrm{NaCl}) g\mathbf{e}_z), &
\end{align}

where $\tilde k(p_c) = k_\mathrm{rw}(S_\mathrm{w}(p_c))$ and $\tilde f(p_c) = f(S_\mathrm{w}(p_c))$.

For convenience and to identify how the different parameters influence the behavior of the model, we will recast the equations in non-dimensional form.

\subsection{Non-dimensional model}\label{sec:lsa_nondim}

We non-dimensionalize the variables by choosing suitable reference values, denoted with subscript ref. Non-dimensional variables are denoted by a hat, and at the same time we simplify some of the notation:

\begin{align}
	\hat{\mathbf x} = \frac{\mathbf x}{\ell_{\text{ref}}},\quad \hat t = \frac{t}{t_{\text{ref}}},\quad \hat P = \frac{p_c}{p_{\text{ref}}}, \quad \hat{\mathbf Q} = \frac{\mathbf{Q}_\mathrm{w}}{Q_\text{ref}}, \quad  \hat{\mathsf{x}}  = \frac{\mathsf{x}_\mathrm{w}^\mathrm{NaCl}}{\mathsf{x}_\text{ref}},\quad \hat\rho = \frac{\rho_\mathrm{w}}{\rho_{\text{ref}}}, \quad \hat\mu = \frac{\mu_\mathrm{w}}{\mu_{\text{ref}}}.    
\end{align}

Note that the salt mole fraction $\mathsf{x}_\mathrm{w}^\mathrm{NaCl}$ is already non-dimensional but is scaled anyway for convenience. The choices for the reference values are summarized in Table \ref{tab:refvalues}. For length reference $\ell_{\mathrm{ref}}$ we choose the domain height $H$. Since we are interested in evaporation, we use the natural time scale for evaporative transport through the domain and hence set $t_\mathrm{ref}=\phi H/E$. For pressure we use the initial hydrostatic pressure, hence $p_\mathrm{ref}=H\rho_\mathrm{ref}g$, while for flow we use the grativational velocity $Q_\mathrm{ref}=K\rho_\mathrm{ref}g/\mu_\mathrm{ref}$. Finally, for the salt concentration, density and viscosity we choose the reference values to be the corresponding initial values, hence $\mathsf{x}_\mathrm{ref}=\mathsf{x}_0^\mathrm{NaCl},\rho_\mathrm{ref} = \rho_\mathrm{w}(\mathsf{x}_0^\mathrm{NaCl}), \mu_\mathrm{ref}= \mu_\mathrm{w}(\mathsf{x}_0^\mathrm{NaCl})$. 
\begin{table} [H]
	\begin{center}
		\begin{tabular}{lll}
			Symbol       & Definition  & Dimension  \\
			\hline
			$\ell_\mathrm{ref}$ & $H$ &  m \\
			$t_\mathrm{ref}$ & $\phi H/E$ & s \\
			$p_\mathrm{ref}$ & $H\rho_\mathrm{ref} g$ &  kg m$^{-1}$ s$^{-2}$\\
			$Q_\mathrm{ref}$ & $K\rho_\mathrm{ref}g/\mu_\mathrm{ref}$ & m s$^{-1}$ \\
			$\mathsf{x}_\mathrm{ref}$ & $\mathsf{x}_0^\mathrm{NaCl}$ & - \\
			$\rho_\mathrm{ref}$ & $\rho_\mathrm{w}(\mathsf{x}_0^\mathrm{NaCl})$ & kg m$^{-3}$ \\
			$\mu_\mathrm{ref}$ & $\mu_\mathrm{w}(\mathsf{x}_0^\mathrm{NaCl})$ & kg m$^{-1}$ s$^{-1}$\\ \hline
		\end{tabular}
		\caption{Overview of reference values used in non-dimensionalization.}
		\label{tab:refvalues}
	\end{center}
\end{table}

Then the non-dimensional model equations are 

\begin{align}
	\partial_{\hat t}(\hat S(\hat P)) + R_E\hat\nabla\cdot\hat{\mathbf Q} &= 0, \label{eq:nondimmass}\\
	\partial_{\hat t}(\hat S(\hat P) \hat{\mathsf{x}}) + \hat\nabla\cdot(R_E\hat{\mathbf Q}\hat{\mathsf x} - \beta\hat f(\hat P)\hat\nabla \hat{\mathsf x}) &= 0, \label{eq:nondimsalt}\\
	\hat{\mathbf Q} = \frac{\hat k(\hat P)}{\hat\mu(\hat{\mathsf{ x}})}(\hat\nabla\hat P-\hat\rho(\hat{\mathsf x})\mathbf{e}_z),& \label{eq:nondimdarcy}
\end{align}

where 

\begin{align}
	R_E = \frac{Q_\mathrm{ref}}{E},\\
	\beta = \frac{\phi^{3/2}D_\mathrm{w}^\mathrm{NaCl}}{EH}.
\end{align}

Here, $R_E$ is the Rayleigh number describing the ratio between the gravitational flow and the flow induced by the evaporative flux at the top boundary. Through the linear stability analysis we will determine a critical Rayleigh number $R_c$, such that instabilities can occur only when $R_c<R_E$. Note that a large value of $R_E$ corresponds to a weaker evaporation compared to gravitational flow. The parameter $\beta$ quantifies the strength of diffusion compared to the evaporation over the height of the porous column. Hence, a larger value of $\beta$ means that the system is more dominated by diffusion.

\begin{remark}
	Note that the Rayleigh number usually depends on the diffusion coefficient and a typical density difference in the system. As we do not have a typical density difference, we use the initial density as reference density, which appears in $Q_\text{ref}$. The Rayleigh number does not explicitly depend on the diffusion coefficient. However, as we will see towards the end of this section, the onset of instabilities will depend $\beta$, which gives an implicit dependence between the critical Rayleigh number and the strength of the diffusion coefficient. 
\end{remark}

The non-dimensional relations in \eqref{eq:nondimmass}-\eqref{eq:nondimdarcy} are

\begin{align}
	\hat S(\hat P ) &= S_\mathrm{wr} + (1-S_\mathrm{wr})(1+\gamma^n\hat P^n)^{-m}, \label{eq:ShatPhat} \\
	\hat f(\hat P) &= (\hat S(\hat P))^{7/2},\\
	\hat k(\hat P) &= (1+\gamma^n\hat P^n)^{-mL}\Big(1-\big(1-(1+\gamma^n\hat P^n)^{-1}\big)^m\Big)^2,
\end{align}

where $\gamma = \alpha p_\mathrm{ref}$. Hence, three dimensionless numbers appear in the dimensionless model; $R_E,\beta$ and $\gamma$. They characterize the behavior of the system. Their definitions are summarized in Table \ref{tab:nondimsymb}. Furthermore, the non-dimensional $\hat\mu(\hat{\mathsf{x}}),\hat\rho(\hat{\mathsf{x}})$ are the corresponding scaled versions of \eqref{eq:variableviscosity}, \eqref{eq:densityDumux} where the scaled mole fraction is used as variable.

\begin{table} [H]
	\begin{center}
		\begin{tabular}{ll}
			Symbol       & Definition    \\
			\hline
			$R_E$ & $Q_\mathrm{ref}/E$ \\
			$\beta$ & $\phi^{3/2}D_\mathrm{w}^\mathrm{NaCl}/EH$  \\
			$\gamma$ & $\alpha p_\mathrm{ref}$ \\ \hline
		\end{tabular}
		\caption{Overview of dimensionless numbers.}
		\label{tab:nondimsymb}
	\end{center}
\end{table}

For the non-dimensional variables, we have the initial conditions

\begin{align}
	\hat{\mathsf{x}}|_{\hat t=0} &= 1, \label{eq:nondimICsalt}\\
	\hat P(\hat z)|_{\hat t=0} &= \hat P_B+\hat z, \label{eq:nondimICp}
\end{align}

where $\hat P_B = p_B/p_\mathrm{ref}$. Note that this initial pressure corresponds to an initial saturation through \eqref{eq:ShatPhat}. 
At the top boundary we apply the boundary conditions

\begin{align}
	\hat{\mathbf Q}|_{\hat z=1} &= \frac{1}{R_E}\mathbf{e}_z, \label{eq:nondimBCqtop}\\
	(\hat{\mathsf{ x}} - \beta\hat f(\hat P)\partial_{\hat z}\hat{\mathsf{x}})|_{\hat z=1} &= 0, \label{eq:nondimBCsalttop}
\end{align}

while for the bottom boundary we apply

\begin{align}
	\hat P|_{\hat z=0} &= \hat P_B, \label{eq:nondimBCpbot}\\
	\hat{\mathsf x}|_{\hat z=0} &= 1. \label{eq:nondimBCsaltbot}
\end{align}

In case of sidewalls we impose

\begin{align}
	\hat{\mathbf Q}|_{\hat x,\hat y=0,\hat \ell}\cdot\mathbf{e}_{x,y} &=0, \label{eq:nondimBCqside}\\
	\partial_{\hat x,\hat y}\hat{\mathsf x}|_{\hat x,\hat y=0,\hat \ell} &= 0, \label{eq:nondimBCsaltside}
\end{align}

where $\hat \ell = \ell/H$ is the non-dimensional width. 
Before continuing to find the ground state it is convenient to restate Darcy's law in terms of the Kirchhoff potential. We then have

\begin{equation}
	\hat{\mathbf Q} = \frac{1}{\hat\mu(\hat{\mathsf x})}(\hat\nabla\Psi - \check{k}(\Psi)\hat\rho(\hat{\mathsf x})\mathbf{e}_z),\label{eq:darcypsi}
\end{equation}

where $\check{k}(\Psi) = \hat k(\hat P(\Psi))$ and

\begin{equation}
	\Psi = \Psi(\hat P) = \int_0^{\hat P} \hat k(\xi)d\xi 
\end{equation}

is the Kirchhoff potential. The Kirchhoff potential is non-dimensional but is for convenience written without hat as it will only be considered in its dimensionless form. 
The function $\Psi(\hat P)$ is monotonically increasing and can therefore be inverted to define

\begin{equation}
	\hat P = \hat P(\Psi). \label{eq:PPsi}
\end{equation}

Hence, we can express the relative permeability, saturation and diffusion in terms of the Kirchhoff potential $\Psi$. We therefore define $\check{S}(\Psi) = \hat S(\hat P(\Psi))$ and $\check{f}(\Psi) = \hat f(\hat P(\Psi))$.

\subsection{Ground state}\label{sec:lsa_groundstate}
We will investigate the solution of the system \eqref{eq:nondimmass}-\eqref{eq:nondimdarcy} under the conditions \eqref{eq:nondimICsalt}-\eqref{eq:nondimBCsaltbot}. In the analysis we use \eqref{eq:darcypsi} instead of \eqref{eq:nondimdarcy}. 
This solution depends on vertical coordinate $\hat z$ and time $\hat t$ only and hence does not include any horizontal variability. Therefore the boundary conditions at the sidewalls \eqref{eq:nondimBCqside}, \eqref{eq:nondimBCsaltside} are satisfied. Since only vertical variability is included, only the vertical velocity component is needed. This solution is the ground state solution and is denoted $\{\hat{W}^0,\hat{\mathsf{x}}^0,\Psi^0\}$, but can equivalently be expressed in terms of $\{\hat{W}^0,\hat{\mathsf{x}}^0,\hat P^0\}$ by using \eqref{eq:PPsi}. Here, $\hat W^0$ is the velocity component of the ground state discharge.

The ground state hence solves the system 

\begin{align}
	\partial_{\hat t}(\hat S(\hat P^0)) + R_E\partial_{\hat z} \hat W^0 &= 0,\label{eq:groundstateS}\\
	\partial_{\hat t}(\hat S(\hat P^0)\hat{\mathsf{x}}^0) + \partial_{\hat z}(R_E\hat W^0\hat{\mathsf{x}}^0-\beta\hat f(\hat P^0)\partial_{\hat z}\hat{\mathsf{x}}^0) &= 0, \label{eq:groundstatesalt}\\
	\hat W^0 = \frac{\hat k(\hat P^0)}{\hat \mu(\hat{\mathsf x}^0)}(\partial_{\hat z}\hat P^0-\hat\rho(\hat{\mathsf x}^0))& \label{eq:groundstateW}
\end{align}

with conditions \eqref{eq:nondimICsalt}-\eqref{eq:nondimBCsaltbot}. Due to the non-linear and coupled nature of these model equations, no explicit forms for the solution is known. We use a Chebyshev-Galerkin collocation method in space and implicit Euler in time to numerically find an approximate solution. 

As $\hat t\to\infty$, the ground state reaches a steady-state denoted $\{\hat{W}^\infty,\hat{\mathsf{x}}^\infty,\Psi^\infty\}$ or $\{\hat{W}^\infty,\hat{\mathsf{x}}^\infty,\hat P^\infty\}$ depending on which form is used. The steady-state ground state now only depends on vertical coordinate $\hat z$ and is given by

\begin{align}
	\hat W^\infty &= \frac{1}{R_E}, \\
	\hat{\mathsf x}^\infty(\hat z) &= \exp{\int_0^{\hat z} \frac{1}{\beta \hat f(\hat P^\infty(\xi))}d\xi},\\
	\hat P^\infty(\hat z) &\text{ solves } \frac{\hat k(\hat P^\infty)}{\hat\mu(\hat{\mathsf{ x}}^\infty)}(\frac{d\hat P^\infty}{d\hat z} -\hat\rho(\hat{\mathsf x}^\infty )) = \frac{1}{R_E},
\end{align}

where simpler closed-form solutions can be found when linear or constant relationships are used for diffusion scaling, relative permeability, viscosity and density.

\begin{remark}
	Note that the model equations have a steady-state solution since the domain is bounded in the vertical direction. In case of an unbounded domain, see for instance \cite{BringedalSatInstab}, the ground state grow unboundedly as $\hat t\to\infty$.
\end{remark}

\subsubsection{Ground state behavior}
\begin{figure}[h!]
	\centering
	\includegraphics[width=0.49\textwidth]{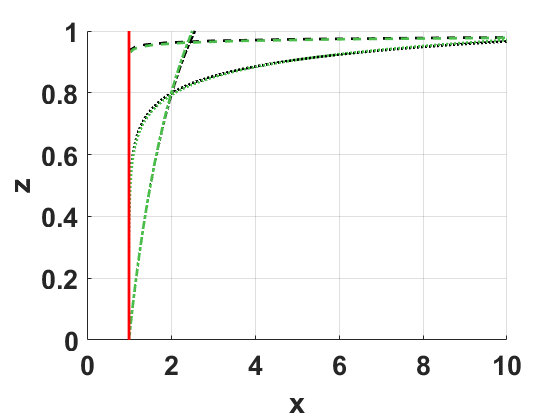}
	\includegraphics[width=0.49\textwidth]{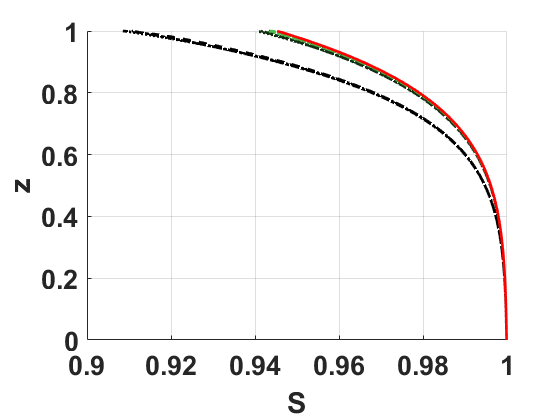}
	\includegraphics[width=0.49\textwidth]{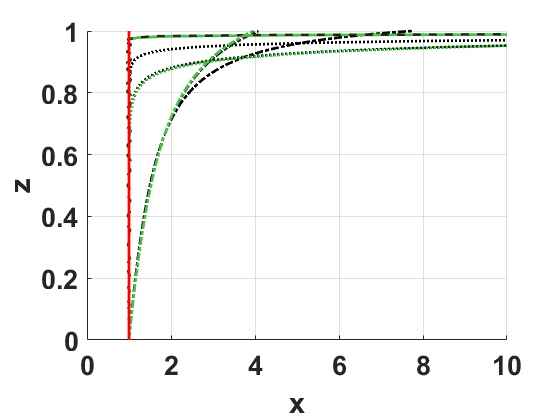}
	\includegraphics[width=0.49\textwidth]{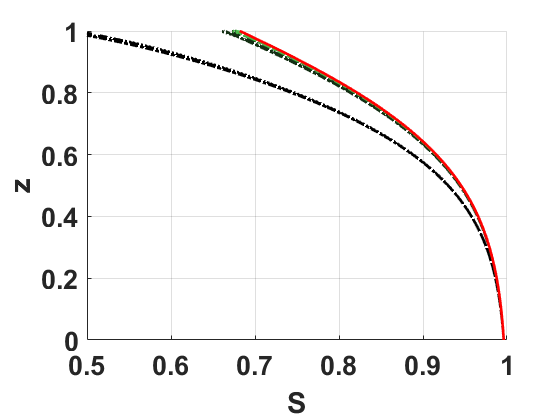}
	\caption{Salt concentration profile (left) and saturation profile (right) using a (non-dimensional) bottom pressure of $\hat P_B = 1.25$ (top row) and $\hat P_B=1.75$ (bottom row). Initial profile in red, while lines corresponding to dashed, dotted and dash-dotted lines are for $\beta=0.01$, $\beta=0.1$ and $\beta=1$, respectively. Black lines correspond to $R_E=10$, while increasingly brighter green correspond to $R_E=10^2,10^3,10^4,10^5$. All lines but the initial red are at non-dimensional time unit $\hat t=1$. Note that many of the curves lie on top of each other.}
	\label{fig:all_GS_pbot125}
\end{figure}

Typical solutions of the ground state using realistic values of $\beta$ and $R_E$, and using properties of the porous medium given in Section \ref{sec:results}, are found in Figure \ref{fig:all_GS_pbot125}. Since the initial saturation profile  is influenced by the choice of bottom pressure, we show solutions using two different bottom pressures as boundary condition. As seen in Figure \ref{fig:all_GS_pbot125}, the time evolution of the saturation profile is very small, except for very low values of $R_E$ (corresponding to e.g.~low permeability or strong evaporation), which decreases the saturation slightly with time. 

The salt concentration profile is mainly influenced by the value of $\beta$ and not of $R_E$. Low values of $\beta$ (corresponding to low diffusion) result in large increases in the concentration near the top of the domain, but less further down. Larger values of $\beta$ give smoother concentration profiles, where increases in salt concentration are more moderate near the top, but where increases can be found also further down in the domain. The salt concentration profile is also influenced by the bottom pressure, and a larger bottom pressure (corresponding to lower saturation) causes the concentration to build up more near the top of the domain, while limiting the diffusion towards the lower parts of the domain. 

The plots in Figure \ref{fig:all_GS_pbot125} are at $\hat t=1$, which is between 0.7 and 300 days, depending on the choice of reference values. Hence, the changes from initial condition seen in Figure \ref{fig:all_GS_pbot125} are potentially extreme, as we are interested in time scales from hours up to a few days.

\begin{figure}[h!]
	\centering
	\includegraphics[width=0.49\textwidth]{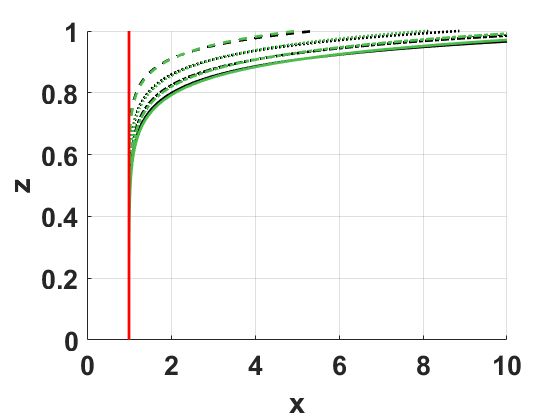}
	\includegraphics[width=0.49\textwidth]{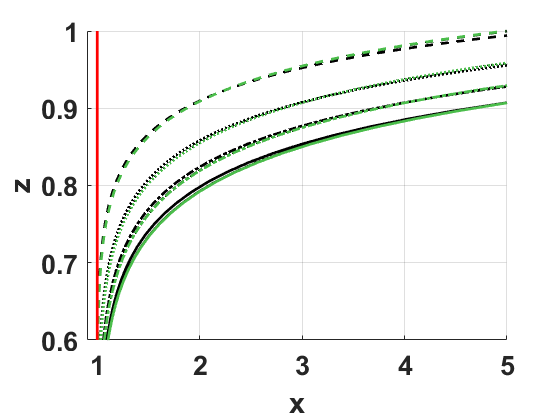}
	\includegraphics[width=0.49\textwidth]{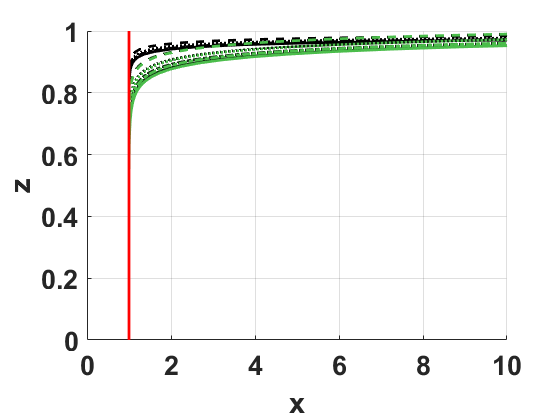}
	\includegraphics[width=0.49\textwidth]{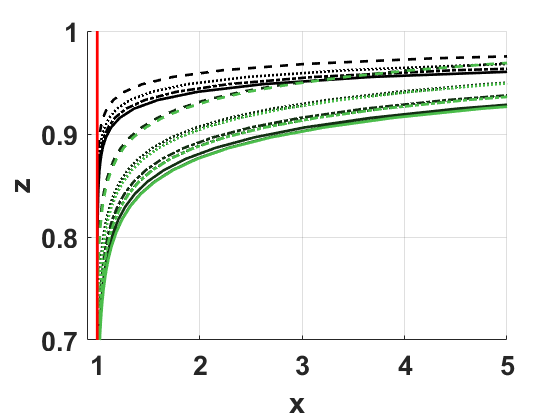}
	\caption{Salt concentration profile (left) and zoomed-in version (right) using a bottom pressure of $\hat P_B = 1.25$ (top row) and $\hat P_B=1.75$ (bottom row). Initial profile in red, while lines corresponding to dashed, dotted, dash-dotted and solid lines are for $\hat t=0.25, 0.5, 0.75, 1$, respectively. Black lines correspond to $R_E=10$, while increasingly brighter green correspond to $R_E=10^2,10^3,10^4,10^5$. Especially for the top row, the darker green lines are barely visible as the brighter green are on top of them. All lines correspond to $\beta=0.1$.}
	\label{fig:Call_GS_pbot125_time}
\end{figure}

In Figure \ref{fig:Call_GS_pbot125_time}, several time steps of the salt concentration profiles for one value of $\beta$ are shown. In all cases, the salt concentration increases with time at the top of the domain, while diffusion causes a gradual increase further down. Especially for the lower bottom pressure, the profiles can be seen to have a small dependence on $R_E$, where low values of $R_E$ give a slightly steeper profile with even larger concentrations near the top of the domain and with less increase further down. For the larger bottom pressure, the time evolution of the case corresponding to the lowest $R_E$ deviates from the other cases, where this case has a very steep profile with very large increase in the salt concentration near the top of the domain, which is to a very little extent propagating downwards in the domain with time. The other cases, corresponding to larger $R_E$, follow the same pattern as for the lower bottom pressure and are barely influenced by the value of $R_E$.

\subsubsection{Simplified ground state}\label{sec:lsa_groundstate_simple}
Motivated by the small changes in saturation observed in Figure \ref{fig:all_GS_pbot125}, we derive a simplified ground state. We take advantage of the fact that the ground state saturation, and hence pressure and Kirchhoff potential and vertical velocity, do not change much with time for most realistic parameter choices. Hence, for the ground state flow potential, pressure and saturation we can use the profile corresponding to the initial profile from \eqref{eq:nondimICp}, while for vertical velocity we use the constant value corresponding to the boundary condition \eqref{eq:nondimBCqtop}, which corresponds to the steady-state ground state $\hat W^\infty$. Hence, only the salt concentration $\hat{\mathsf x}^0$ have to be solved for, and \eqref{eq:groundstatesalt} can be replaced with the simpler

\begin{equation}
	S^0\partial_{\hat t}\hat{\mathsf x}^0 + \partial_{\hat z}(\hat{\mathsf x}^0-\beta \hat f(S^0)\partial_{\hat z}\hat{\mathsf x}^0) = 0, \label{eq:groundstatesaltsimple}
\end{equation}

and otherwise with same initial \eqref{eq:nondimICsalt} and boundary conditions \eqref{eq:nondimBCsaltbot},\eqref{eq:nondimBCsalttop} as before. In \eqref{eq:groundstatesaltsimple}, $S^0$ is the saturation corresponding to the initial saturation. Note that this simplified ground state does not depend explicitly on $R_E$ anymore.

\begin{figure}[h!]
	\centering
	\includegraphics[width=0.49\textwidth]{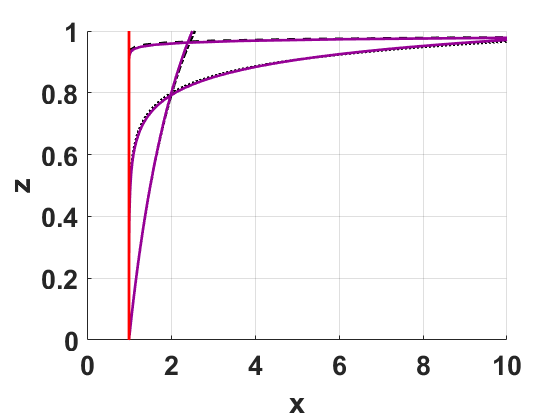}
	\includegraphics[width=0.49\textwidth]{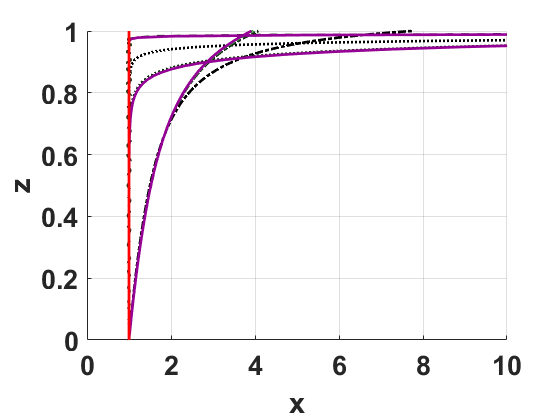}
	\caption{Salt concentration profile, repetition of the left part of Figure \ref{fig:all_GS_pbot125}, with the simplified salt concentration overlayed with purple lines. Here, the left figure corresponds to $\hat P_B=1.25$ and the right figure to $\hat P_B = 1.75$. Lines corresponding to dashed, dotted and dash-dotted lines are for $\beta=0.01$, $\beta=0.1$ and $\beta=1$, respectively. All lines but the initial red are at non-dimensional time $\hat t=1$.}
	\label{fig:call_GS_pbot125_simple}
\end{figure}

\begin{figure}[h!]
	\centering
	\includegraphics[width=0.49\textwidth]{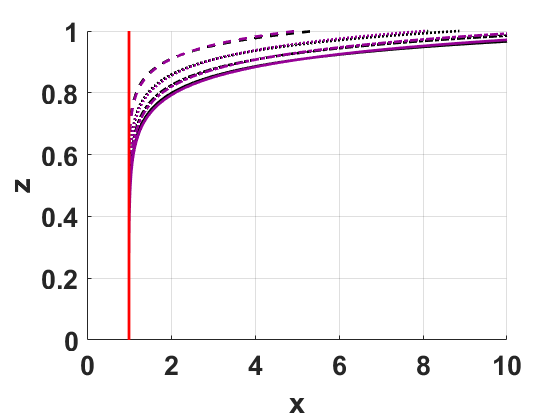}
	\includegraphics[width=0.49\textwidth]{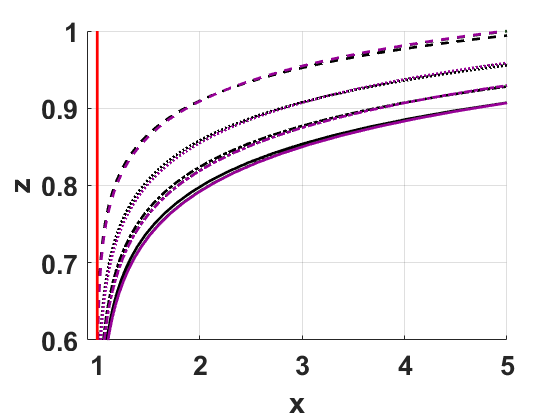}
	\includegraphics[width=0.49\textwidth]{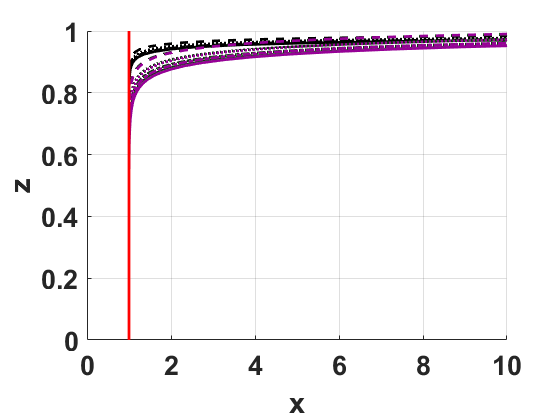}
	\includegraphics[width=0.49\textwidth]{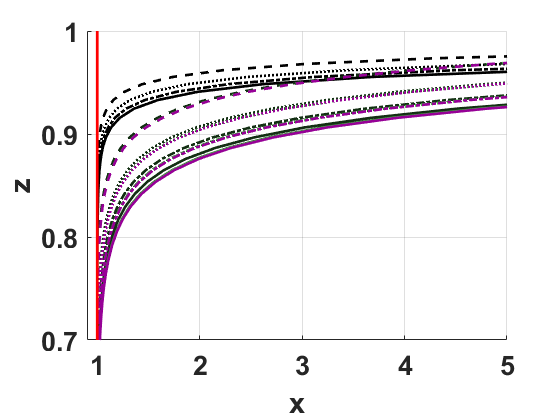}
	\caption{Salt concentration profile (left) and zoomed-in version (right) using a bottom pressure of $\hat P_B = 1.25$ (top row) and $\hat P_B=1.75$ (bottom row). Repetition of Figure \ref{fig:Call_GS_pbot125_time}, with the simplified salt concentration overlayed with purple lines. Lines corresponding to dashed, dotted, dash-dotted and solid lines are for $\hat t=0.25, 0.5, 0.75, 1$, respectively.}
	\label{fig:Call_GS_pbot125_time_simple}
\end{figure}

Salt concentration profiles from the simplified ground state can be found in Figure \ref{fig:call_GS_pbot125_simple} and \ref{fig:Call_GS_pbot125_time_simple}. The simplified ground state matches overall very well with the original ground state solution, although small deviations can be observed when comparing with low Rayleigh numbers. Note that also the curves corresponding to $R_E=10^2,10^3,10^4,10^5$ are in Figure \ref{fig:call_GS_pbot125_simple} and \ref{fig:Call_GS_pbot125_time_simple}, but these are so close to the purple lines corresponding to the simplified ground state that they cannot be visually separated. Hence, except for very low Rayleigh numbers ($R_E\approx10$), the simplified ground state appears to be a very good approximation for the salt concentration.


\subsection{Linear perturbation and eigenvalue problem}\label{sec:lsa_eig}
We investigate the linear stability of the ground state $\{\hat{W}^0,\hat{\mathsf{x}}^0,\Psi^0\}$ or $\{\hat{W}^0,\hat{\mathsf{x}}^0,\hat P^0\}$ by perturbing it. We use here the original ground state solution and not the simplified version \eqref{eq:groundstatesaltsimple}. Hence, we consider

\begin{align}
	\hat{\mathbf Q}(\hat t,\hat x,\hat y,\hat z) &= \hat W^0(\hat t,\hat z)\mathbf e_z + {\mathbf q}(\hat t,\hat x,\hat y,\hat z),\\
	\hat{\mathsf x}(\hat t,\hat x,\hat y,\hat z) &= \hat{\mathsf x}^0(\hat t,\hat z) + \chi(\hat t,\hat x,\hat y,\hat z), \\
	\hat P(\hat t,\hat x,\hat y,\hat z) &= \hat P^0(\hat t,\hat z) + p(\hat t,\hat x,\hat y,\hat z),
\end{align}

or equivalently, by replacing the last equation by

\begin{equation}
	\Psi(\hat t,\hat x,\hat y,\hat z) = \Psi^0(\hat t,\hat z) + \psi(\hat t,\hat x,\hat y,\hat z),
\end{equation}

where $\psi = \int_{\hat P^0}^{\hat P^0+p}\hat k(\xi)d\xi$. Here, $\mathbf q, \chi, p$ and $\psi$ are perturbed velocity, salt concentration, pressure and Kirchhoff potential, respectively, and are all assumed to be small quantities. They are all non-dimensional, but are for convenience written without hat. We obtain equations and boundary conditions for the perturbed variables by inserting them into \eqref{eq:nondimmass}-\eqref{eq:nondimdarcy} and \eqref{eq:nondimBCqtop}-\eqref{eq:nondimBCsaltside}. Initial conditions for the perturbed variables are handled in a different manner and will be explained further below. Hence, the perturbed variables fulfill

\begin{align}
	\partial_{\hat t}(\check{S}(\Psi^0+\psi)) + R_E\hat\nabla\cdot(\hat W^0\mathbf{e}_z + \mathbf{q}) &= 0, \label{eq:perturbS0}\\
	\partial_{\hat t}(\check{S}(\Psi^0+\psi)(\hat{\mathsf{x}}^0+\chi))
	+ \hat\nabla\cdot(R_E(\hat W^0\mathbf{e}_z+\mathbf{q})(\hat{\mathsf{x}}^0+\chi) - \beta\check{f}(\Psi^0+\psi)\hat\nabla(\hat{\mathsf{x}}^0+\chi)) &= 0, \label{eq:perturbX0}\\
	\hat W^0\mathbf{e}_z + \mathbf{q} = \frac{1}{\hat\mu(\hat{\mathsf x}^0+\chi)}\hat\nabla(\Psi^0+\psi) - \check{k}(\Psi^0+\psi)\frac{\hat\rho(\hat{\mathsf x}^0+\chi)}{\hat\mu(\hat{\mathsf x}^0+\chi)}\mathbf e_z.& \label{eq:perturbW0}
\end{align}

Since $\hat{\mathbf Q},\hat{\mathsf x}, \hat P$ fulfill the same boundary conditions as the ground state, the perturbed variables need to fulfill corresponding homogeneous boundary conditions at the top and bottom boundaries. Hence, we find

\begin{align}
	\mathbf q|_{\hat z=1}\cdot\mathbf e_z &= 0,\\
	(\hat{\mathsf x}^0 + \chi -\beta\check{f}(\Psi^0+\psi)\partial_{\hat z}(\hat{\mathsf x}^0+\chi))|_{\hat z=1} &= 0 \label{eq:nonlinBCperturbsalt}
\end{align}

for the top boundary, and

\begin{align}
	p|_{\hat z=0} &= 0, \label{eq:perturbBCpbot}\\
	\chi|_{\hat z=0} &= 0
\end{align}

at the bottom boundary, and

\begin{align}
	\mathbf q|_{\hat x,\hat y=0,\ell}\cdot\mathbf e_{x,y} &= 0, \\
	\partial_{\hat x,\hat y}\chi|_{\hat x,\hat y=0,\ell} &= 0 \label{eq:perturbsidebcchi}
\end{align}

on any sidewalls.

In the following we take advantage of the fact that the perturbed variables are small and we therefore linearize the above equations. First of all, the boundary condition \eqref{eq:nonlinBCperturbsalt} is replaced with

\begin{equation}
	(\chi - \beta\check{f}(\Psi^0)\partial_{\hat z}\chi - \beta\check{f}'(\Psi^0)\partial_{\hat z}\hat{\mathsf x}^0\psi)|_{\hat z=1} =0
\end{equation}

by expanding $\check{f}(\Psi^0+\psi)$ around $\Psi^0$ and removing terms including both $\chi$ and $\psi$ as they are assumed to be small. In a similar fashion, \eqref{eq:perturbS0} becomes

\begin{equation}
	\partial_{\hat t}(\check{S}'(\Psi^0)\psi) + R_E\hat\nabla\cdot\mathbf q =0, \label{eq:perturbpsi}
\end{equation}

while \eqref{eq:perturbX0} is linearized to

\begin{align}
	\check{S}(\Psi^0)\partial_{\hat t}\chi &+ R_E\hat W^0\partial_{\hat z}\chi -\beta\hat\nabla\cdot(\check{f}(\Psi^0)\hat\nabla\chi) \nonumber \\ &= -R_E w\partial_{\hat z}\hat{\mathsf x}^0 - \check{S}'(\Psi^0)\partial_{\hat t}\hat{\mathsf x}^0\psi +\beta\partial_{\hat z}(\check{f}'(\Psi^0)\partial_{\hat z}\hat{\mathsf x}^0)\psi + \beta\check{f}'(\Psi^0)\partial_{\hat z}\hat{\mathsf x}^0\partial_{\hat z}\psi, \label{eq:perturbchi}
\end{align}

where $w$ is the vertical component of the perturbed velocity $\mathbf q$. Finally, \eqref{eq:perturbW0} is linearized to

\begin{align}
	\mathbf q = \frac{1}{\hat\mu(\hat{\mathsf x}^0)}\hat\nabla\psi - \frac{\hat\mu'(\hat{\mathsf x}^0)\chi}{(\hat\mu(\hat{\mathsf x}^0))^2}\hat\nabla\Psi^0 
	- \Big(\check{k}(\Psi^0)\frac{\hat\rho'(\hat{\mathsf x}^0)}{\hat\mu(\hat{\mathsf x}^0)}\chi + \check{k}'(\Psi^0)\frac{\hat\rho(\hat{\mathsf x}^0)}{\hat\mu(\hat{\mathsf x}^0)}\psi - \check{k}(\Psi^0)\frac{\hat\rho(\hat{\mathsf x}^0)\hat\mu'(\hat{\mathsf x}^0)}{(\hat\mu(\hat{\mathsf x}^0))^2}\chi \Big)\mathbf e_z \label{eq:perturbq}
\end{align}

We want to separate the vertical and horizontal variability. First note from \eqref{eq:perturbq} that

\begin{align}
	w = \frac{1}{\hat\mu(\hat{\mathsf x}^0)}\partial_{\hat z}\psi - \frac{\hat\mu'(\hat{\mathsf x}^0)\chi}{(\hat\mu(\hat{\mathsf x}^0))^2}\partial_{\hat z}\Psi^0 
	- \check{k}(\Psi^0)\frac{\hat\rho'(\hat{\mathsf x}^0)}{\hat\mu(\hat{\mathsf x}^0)}\chi - \check{k}'(\Psi^0)\frac{\hat\rho(\hat{\mathsf x}^0)}{\hat\mu(\hat{\mathsf x}^0)}\psi + \check{k}(\Psi^0)\frac{\hat\rho(\hat{\mathsf x}^0)\hat\mu'(\hat{\mathsf x}^0)}{(\hat\mu(\hat{\mathsf x}^0))^2}\chi \label{eq:perturbw}
\end{align}

while we have from taking the divergence of \eqref{eq:perturbq} that

\begin{equation}
	\hat\nabla\cdot\mathbf q = \partial_{\hat z}w + \frac{1}{\hat\mu(\hat{\mathsf x}^0)}\hat\nabla^2_h\psi, 
\end{equation}

where $\hat\nabla^2_h$ is the horizontal Laplace operator. Hence, \eqref{eq:perturbpsi} can be written

\begin{equation}
	\partial_{\hat t}(\check{S}'(\Psi^0)\psi) + R_E\partial_{\hat z} w + \frac{R_E}{\hat\mu(\hat{\mathsf x}^0)}\hat\nabla_h^2\psi = 0. \label{eq:perturbpsi2}
\end{equation}

By differentiating \eqref{eq:perturbpsi2} with respect to $\hat z$ and inserting the horizontal Laplacian of \eqref{eq:perturbw} to rewrite the last term of \eqref{eq:perturbpsi2}, we obtain

\begin{align}
	\hat\nabla^2w &+ \frac{1}{R_E}\partial_{\hat z\hat t}(\check{S}'(\Psi^0)\psi) = \Big(\frac{\hat\mu'(\hat{\mathsf x}^0)}{(\hat\mu(\hat{\mathsf x}^0))^2}\partial_{\hat z}\hat{\mathsf x}^0 - \check{k}'(\Psi^0)\frac{\hat\rho(\hat{\mathsf x}^0)}{\hat\mu(\hat{\mathsf x}^0)}\Big)\hat\nabla_h^2\psi\nonumber\\  &+ \Big(\frac{\hat\rho(\hat{\mathsf x}^0)\hat\mu'(\hat{\mathsf x}^0)}{(\hat\mu(\hat{\mathsf x}^0))^2}\check{k}(\Psi^0)-\frac{\hat\rho'(\hat{\mathsf x}^0)}{\hat\mu(\hat{\mathsf x}^0)}\check{k}(\Psi^0)-\frac{\hat\mu'(\hat{\mathsf x}^0)}{(\hat\mu(\hat{\mathsf x}^0))^2}\partial_{\hat z}\Psi^0\Big)\hat\nabla_h^2\chi \label{eq:perturbpsi3}
\end{align}

Since we have a linear boundary value problem for the perturbed variables, where no coefficients depend on the horizontal spatial coordinates, we consider solutions of the form

\begin{equation}
	\{w,\chi,\psi\}(\hat t,\hat x,\hat y,\hat z) = \{\tilde w,\tilde\chi,\tilde\psi\}(\hat t,\hat z)\cos(\hat a_x\hat x)\cos(\hat a_y\hat y),
\end{equation}

where $\tilde w,\tilde \chi,\tilde\psi$ are amplitudes and $\hat a_x,\hat a_y$ are horizontal, angular wavenumbers. In the general case without influence of any sidewalls we have that $\hat a_x$ and $\hat a_y$ can be any positive number. For a two-dimensional bounded domain, one would set $\hat a_y=0$ and restrict $\hat a_x$ such that boundary conditions on the sidewalls are fulfilled. From \eqref{eq:perturbsidebcchi} we get that

\begin{equation}
	\hat a_x = n_x\frac{2\pi}{\ell},\quad n_x=1,2,\dots \label{eq:arest}
\end{equation}

where $n_x$ is the number of waves in the domain in the $x$ direction, and one would have similarly for $y$ direction. 
We introduce $\hat a^2=\hat a_x^2+\hat a_y^2$. If $\hat a_y=0$, then is $\hat a=\hat a_x$. 
Using this horizontal variability, we get that \eqref{eq:perturbchi} is now

\begin{align}
	\check{S}(\Psi^0)\partial_{\hat t}\tilde \chi &+ R_E\hat W^0\partial_{\hat z}\tilde \chi -\beta\partial_{\hat z}(\check{f}(\Psi^0))\partial_{\hat z}\tilde \chi - \beta\check{f}(\Psi^0)\partial_{\hat z}^2\tilde \chi + \beta \hat a^2\check{f}(\Psi^0)\tilde \chi \nonumber \\ &= -R_E \tilde w\partial_{\hat z}\hat{\mathsf x}^0 - \check{S}'(\Psi^0)\partial_{\hat t}\hat{\mathsf x}^0\tilde \psi +\beta\partial_{\hat z}(\check{f}'(\Psi^0)\partial_{\hat z}\hat{\mathsf x}^0)\tilde \psi + \beta\check{f}'(\Psi^0)\partial_{\hat z}\hat{\mathsf x}^0\partial_{\hat z}\tilde \psi, \label{eq:perturbchi2}
\end{align}

and from \eqref{eq:perturbpsi2} we get

\begin{equation}
	\frac{1}{R_E}\check{S}'(\Psi^0)\partial_{\hat t}\tilde \psi + \partial_{\hat z}\tilde w = A\tilde \psi, \label{eq:perturbpsi4}
\end{equation}

where

\begin{equation}
	A = A(\hat t,\hat z) = \frac{\hat a^2}{\hat\mu(\hat{\mathsf x}^0)} - \frac{1}{R_E}\partial_{\hat t}(\check{S}'(\Psi^0)). \label{eq:Aeq}
\end{equation}

Note that the boundary condition \eqref{eq:perturbBCpbot} corresponds to $\tilde \psi|_{\hat z=0}=0$, which through \eqref{eq:perturbpsi4} becomes

\begin{equation}
	\partial_{\hat z}\tilde w|_{\hat z=0} =0.
\end{equation}

By using the product rule on $\partial_{\hat z\hat t}(\check{S}'(\Psi^0)\tilde \psi)$ in \eqref{eq:perturbpsi3}, rewriting the horizontal variability with \eqref{eq:arest}, and using \eqref{eq:perturbw} as well as \eqref{eq:Aeq} to simplify terms, we obtain that \eqref{eq:perturbpsi3} can be written

\begin{align}
	\partial_{\hat z}^2\tilde w &-A\hat\mu(\hat{\mathsf x}^0)\tilde w + \frac{1}{R_E}(\partial_{\hat z}(\check{S}'(\Psi^0))\partial_{\hat t}\tilde \psi + \check{S}'(\Psi^0)\partial_{\hat z\hat t}\tilde \psi ) - \partial_{\hat z}A\tilde \psi -A\check{k}'(\Psi^0)\hat\rho(\hat{\mathsf x}^0)\tilde \psi \nonumber\\  &= A\Big(\check{k}(\Psi^0)(\hat\rho'(\hat{\mathsf x}^0)-\frac{\hat\rho(\hat{\mathsf x}^0)\hat\mu'(\hat{\mathsf x}^0)}{\hat\mu(\hat{\mathsf x}^0)})+\frac{\hat\mu'(\hat{\mathsf x}^0)}{\hat\mu(\hat{\mathsf x}^0)}\partial_{\hat z}\Psi^0\Big)\tilde \chi.  \label{eq:perturbpsi5}
\end{align}

The governing equations are hence \eqref{eq:perturbchi2}, \eqref{eq:perturbpsi4} and \eqref{eq:perturbpsi5}, for the unknowns $\tilde \chi, \tilde w$ and $\tilde \psi$. Using the linearity of \eqref{eq:perturbchi2}, \eqref{eq:perturbpsi4} and \eqref{eq:perturbpsi5}, and assuming that the coefficients in these equations do not depend on time $\hat t$, we can further use that

\begin{equation}
	\{\tilde w,\tilde\chi,\tilde\psi\}(\hat t,\hat z) = \{\hat w,\hat\chi,\hat\psi\}(\hat z) e^{\sigma\hat t} 
\end{equation}

where $\sigma$ is the exponential growth rate in time, and $\hat w,\hat\chi,\hat\psi$ describe vertical variability. However, since the ground state variables $\hat W^0,\Psi^0,\hat{\mathsf x}^0$ depend on time, this does not directly apply. We circumvent this by applying the quasi-steady-state approach. That means, we freeze the ground state at a fixed time $\hat t^\ast$ and investigate only small increments away from the fixed time. Then, $\hat\tau = \hat t-\hat t^\ast$ would be the new time variable, for which we apply 

\begin{equation}
	\{\tilde w,\tilde\chi,\tilde\psi\}(\hat \tau,\hat z) = \{\hat w,\hat\chi,\hat\psi\}(\hat z) e^{\sigma\hat \tau} .
\end{equation}

We have that $\sigma>0$ corresponds to exponential growth of the perturbation, while $\sigma<0$ would give exponential decay of their strength with time. Neutral stability is when $\sigma=0$, which corresponds to the critical point where we have exchange of stability \cite{van2001stability}. Hence, to investigate when there can be perturbations, it suffices to investigate the case where $\sigma=0$, see for example discussion in \cite{BringedalSatInstab}. In this case, time derivatives of the perturbations disappear. From \eqref{eq:perturbpsi4} we get that

\begin{equation}
	\hat w' = A\hat\psi, \label{eq:psieq}
\end{equation}

where derivative of $\hat w$ means derivative with respect to $\hat z$. We can use \eqref{eq:psieq} to eliminate $\hat \psi$ from \eqref{eq:perturbchi2} and \eqref{eq:perturbpsi5}, and we can use the corresponding version of \eqref{eq:perturbw} to eliminate $\partial_{\hat z}\hat \psi$. We hence can formulate two coupled ordinary differential equations for $\hat\chi$ and $\hat w$ only. From \eqref{eq:perturbchi2} we get

\begin{align}
	\beta\check{f}&(\Psi^0)\hat\chi''  + (\beta\partial_{\hat z}(\check{f}(\Psi^0))- R_E\hat W^0)\hat\chi' \nonumber\\ &+ \Big(-\beta\hat a^2\check{f}(\Psi^0)+\beta\check{f}'(\Psi^0)\partial_{\hat z}\hat{\mathsf x}^0\big(\frac{\hat\mu'(\hat{\mathsf x}^0)}{\hat\mu(\hat{\mathsf x}^0)}\partial_{\hat z}\Psi^0+\check{k}(\Psi^0)\hat\rho'(\hat{\mathsf x}^0)-\check{k}(\Psi^0)\hat\rho(\hat{\mathsf x}^0)\frac{\hat\mu'(\hat{\mathsf x}^0)}{\hat\mu(\hat{\mathsf x}^0)}\big)\Big)\hat\chi \nonumber \\ = &\frac{1}{A}(\check{S}'(\Psi^0)\partial_{\hat t}\hat{\mathsf x}^0-\beta\partial_{\hat z}(\check{f}'(\Psi^0)\partial_{\hat z}\hat{\mathsf x}^0)-\beta\check{f}'(\Psi^0)\partial_{\hat z}\hat{\mathsf x}^0\check{k}'(\Psi^0)\hat\rho(\hat{\mathsf x}^0) )\hat w'\nonumber \\ &
	+ (R_E\partial_{\hat z}\hat{\mathsf x}^0-\beta\check{f}'(\Psi^0)\partial_{\hat z}\hat{\mathsf x}^0\hat\mu(\hat{\mathsf x}^0))\hat w \label{eq:chifinal}
\end{align}

while \eqref{eq:perturbpsi5} becomes

\begin{align}
	\hat w'' &- (\frac{\partial_{\hat z}A}{A}+\check{k}'(\Psi^0)\hat\rho(\hat{\mathsf x}^0))\hat w' - A\hat\mu(\hat{\mathsf x}^0)\hat w \nonumber \\ &
	= A\Big( \frac{\hat\mu'(\hat{\mathsf x}^0)}{\hat\mu(\hat{\mathsf x}^0)}\partial_{\hat z}\Psi^0+\check{k}(\Psi^0)\hat\rho'(\hat{\mathsf x}^0)-\check{k}(\Psi^0)\frac{\hat\rho(\hat{\mathsf x}^0)\hat\mu'(\hat{\mathsf x}^0)}{\hat\mu(\hat{\mathsf x}^0)}\Big)\hat \chi. \label{eq:wfinal}
\end{align}

Note that in \eqref{eq:chifinal} and \eqref{eq:wfinal}, the Rayleigh number $R_E$ appears. Given the ground state $\{\hat W^0,\hat{\mathsf x}^0,\Psi^0\}$ evaluated at a fixed time $\hat t^\ast$, we can consider this as an eigenvalue problem in terms of $\{\hat\chi,\hat w\}$. The objective is to determine the smallest positive eigenvalue that allows presence of perturbation. This is done by minimizing over the possible angular wavenumbers $\hat a$ at a fixed time. Hence, we arrive at searching for the critical Rayleigh number $R_c(\hat t^\ast) = \min_{\hat a} R_E(\hat a,\hat t^\ast)$. Hence, we will eventually find a critical wavenumber $R_c(\hat t)$, where we for simplicity use $\hat t$ as notation instead of $\hat t^\ast$.

\subsubsection{Resulting eigenvalue problem}
Given $\hat a>0$ and for fixed $\hat t$, and given the ground state $\{\hat W^0,\hat{\mathsf x}^0,\Psi^0\}$ evaluated at time $\hat t$, find the smallest $R_E(\hat a,\hat t)$ such that

\begin{equation}
	\left.
	\begin{array}{lr}
		\hat w \text{ and } \hat\chi \text{ solves } \eqref{eq:chifinal}, \eqref{eq:wfinal} & 0<\hat z<1,\\
		\text{where } \hat w \text{ and } \hat\chi \text{ fulfill} & \\
		\hat w=0, \hat\chi -\beta\check{f}(\Psi^0)\hat\chi' -\beta\check{f}'(\Psi^0)\frac{1}{A}\partial_{\hat z}\hat{\mathsf x}^0\hat w' = 0 & \hat z=1\\
		\hat w'=0, \hat\chi = 0 & \hat z=0
	\end{array}
	\right\} \label{eq:evp}
\end{equation}

has a non-trivial solution. We want to find the minimum value $R_c(\hat t) :=\min_{\hat a} R_E(\hat a,\hat t)$.

In the case of a horizontally bounded domain, \eqref{eq:arest} applies.

\subsubsection{Simplified eigenvalue problem}
The resulting eigenvalue problem \eqref{eq:evp} is not easy to solve as the eigenvalue $R_E$ appears in both \eqref{eq:chifinal} but also in the ground state \eqref{eq:groundstateS}, \eqref{eq:groundstatesalt} and \eqref{eq:nondimBCqtop}. Since the ground state has no known explicit solution, the ground state needs to be discretized and time-stepped up to time $\hat t$ for a prescribed $R_E$, which is then to be solved for in the eigenvalue problem \eqref{eq:evp}. Although this resulting non-linear system of equations can be solved for $R_E$ by using an iterative solver, a possible approach to circumvent this issue is to rather use the simplified ground state discussed in Section \ref{sec:lsa_groundstate_simple}. Then, ground state saturation (and correspondingly pressure and flow potential) corresponds to the initial profile from \eqref{eq:nondimICp}, the vertical velocity is equal to the value at the boundary condition \eqref{eq:nondimBCqtop}, while salt concentration is given by \eqref{eq:groundstatesaltsimple}. 
Hence, the salt ground state does not depend on the Rayleigh number, neither will the initial potential $\Psi^0$. The vertical velocity $\hat W^0$ is equal to $1/R_E$, but this value can be inserted directly into \eqref{eq:chifinal}. Since $\Psi^0$ is no longer time-dependent, the expression for $A$ \eqref{eq:Aeq} is simplified to

\begin{equation}
	A = A(\hat t,\hat z) = \frac{\hat a^2}{\hat\mu(\hat{\mathsf x}^0)}.
\end{equation}

Hence, we arrive at a simplified eigenvalue problem which is linear in $R_E$ and is therefore easier to solve.

Given $\hat a>0$ and for fixed $\hat t$, and given the salt ground state $\hat{\mathsf x}^0$ evaluated at time $\hat t$ and the initial condition $\Psi^0$, find the smallest $R_E(\hat a,\hat t)$ such that

\begin{equation}
	\left.
	\begin{array}{lr}
		\beta\check{f}(\Psi^0)\chi'' + (\beta\partial_{\hat z}(\check{f}(\Psi^0))-1)\chi' & \\+ \big(-\beta\hat a^2\check{f}(\Psi^0)+\beta\check{f}'(\Psi^0)\partial_{\hat z}\hat{\mathsf x}^0\big( \frac{\hat\mu'(\hat{\mathsf x}^0)}{\hat\mu(\hat{\mathsf x}^0)}\partial_{\hat z}\Psi^0+\check{k}(\Psi^0)\hat\rho'(\hat{\mathsf x}^0)-\check{k}(\Psi^0)\hat\rho(\hat{\mathsf x}^0)\frac{\hat\mu'(\hat{\mathsf x}^0)}{\hat\mu(\hat{\mathsf x}^0)}\big)\big)\hat\chi &  \\
		= \frac{\hat\mu(\hat{\mathsf x}^0)}{\hat a^2}\big(\check{S}'(\Psi^0)\partial_{\hat t}\hat{\mathsf x}^0-\beta\partial_{\hat z}(\check{f}'(\Psi^0)\partial_{\hat z}\hat{\mathsf x}^0)-\beta\check{f}'(\Psi^0)\partial_{\hat z}\hat{\mathsf x}^0\check{k}'(\Psi^0)\hat\rho(\hat{\mathsf x}^0) \big)\hat w' &  \\
		+(R_E\partial_{\hat z}\hat{\mathsf x}^0-\beta\check{f}'(\Psi^0)\partial_{\hat z}\hat{\mathsf x}^0\hat\mu(\hat{\mathsf x}^0))\hat w & \hspace{-4cm} 0<\hat z<1\\ 
		\hat w'' + \big( \frac{\hat\mu'(\hat{\mathsf x}^0)\partial_{\hat z}\hat{\mathsf x}^0}{\hat\mu(\hat{\mathsf x}^0)} - \check{k}'(\Psi^0)\hat\rho(\hat{\mathsf x}^0)\big)\hat w' - \hat a^2\hat w & \\ =\frac{\hat a^2}{\hat\mu(\hat{\mathsf x}^0)}\Big( \frac{\hat\mu'(\hat{\mathsf x}^0)}{\hat\mu(\hat{\mathsf x}^0)}\partial_{\hat z}\Psi^0+\check{k}(\Psi^0)\hat\rho'(\hat{\mathsf x}^0)-\check{k}(\Psi^0)\frac{\hat\rho(\hat{\mathsf x}^0)\hat\mu'(\hat{\mathsf x}^0)}{\hat\mu(\hat{\mathsf x}^0)}\Big)\hat \chi & \hspace{-4cm} 0<\hat z<1\\
		\text{where } \hat w \text{ and } \hat\chi \text{ fulfill} & \\
		\hat w=0, \hat\chi -\beta\check{f}(\Psi^0)\hat\chi' -\beta\check{f}'(\Psi^0)\frac{\hat\mu(\hat{\mathsf x}^0)}{\hat a^2}\partial_{\hat z}\hat{\mathsf x}^0\hat w' = 0 & \hspace{-4cm}\hat z=1\\
		\hat w'=0, \hat\chi = 0 & \hspace{-4cm}\hat z=0
	\end{array}
	\right\} \label{eq:evp_simple}
\end{equation}

has a non-trivial solution. Again we seek the minimum value $R_c(\hat t) :=\min_{\hat a} R_E(\hat a,\hat t)$. We call the corresponding non-trivial solution $\hat{\mathsf{x}}^0$ the eigenprofile of the eigenvalue problem. The eigenprofile corresponds to the shape of the perturbation as instabilities can arise.

In the case of a horizontally bounded domain, \eqref{eq:arest} applies.

The simplified eigenvalue problem \eqref{eq:evp_simple} should give solutions that are good approximations of the solutions of the full eigenvalue problem \eqref{eq:evp} when the simplified ground state is a good approximation of the original ground state. As discussed in Section \ref{sec:lsa_groundstate_simple}, this is the case when the Rayleigh number $R_E$ is not very small, hence we can trust the solutions of the simplified eigenvalue problem for $R_E\gg10$.

\subsection{Solution of the eigenvalue problem}\label{sec:lsa_sol}
The eigenvalue problem \eqref{eq:evp_simple} is solved by discretizing the vertical variability of the equations in \eqref{eq:evp_simple} using a Chebyshev-Galerkin approach. This results in a linear system of equations for the discrete points. Through the eigenvalues of the resulting matrix, we find the required eigenvalue of \eqref{eq:evp_simple}. The corresponding eigenprofile will therefore be the connected eigenvector.

\begin{figure}[h!]
	\centering
	\includegraphics[width=0.49\textwidth]{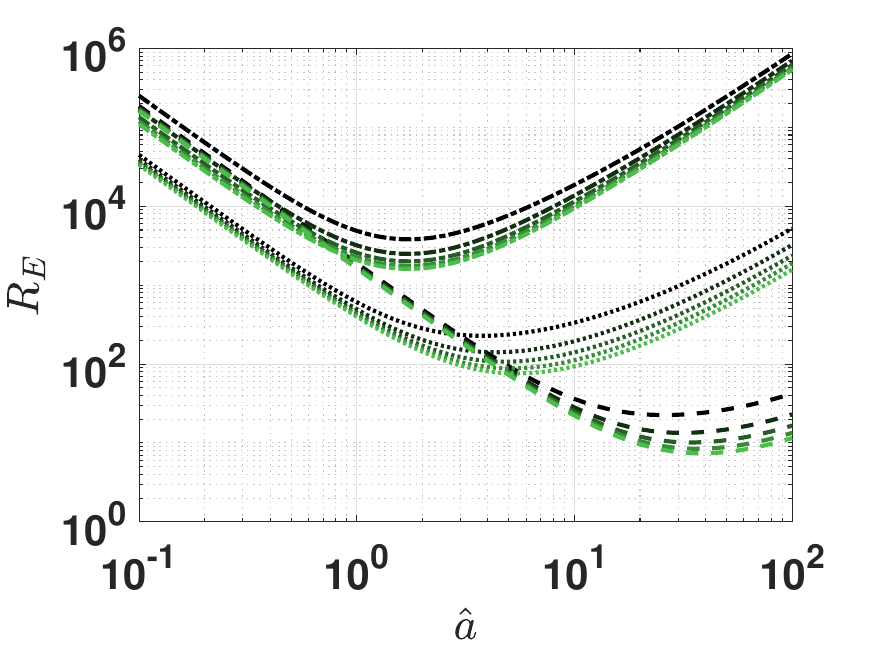}
	\includegraphics[width=0.49\textwidth]{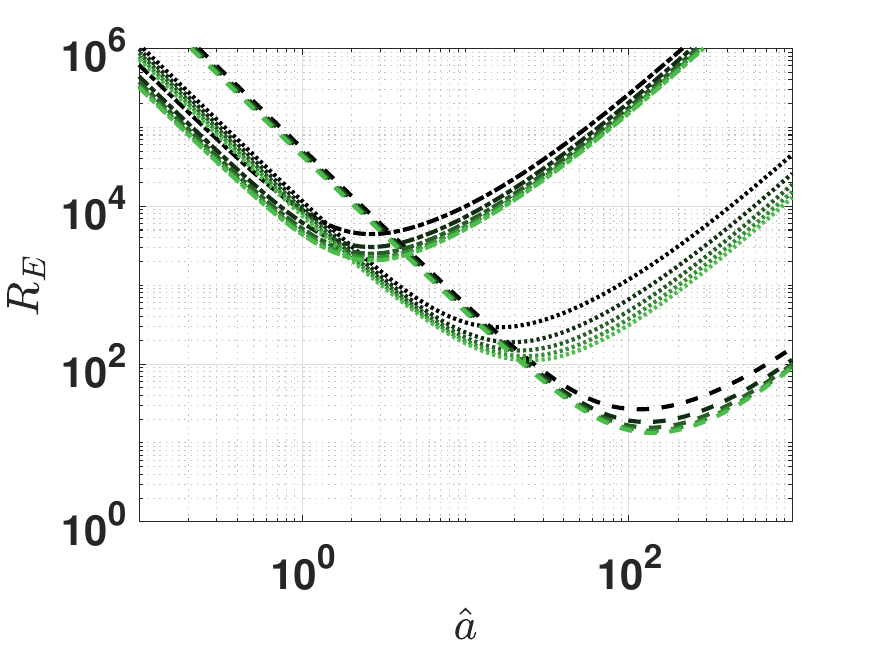}
	\includegraphics[width=0.49\textwidth]{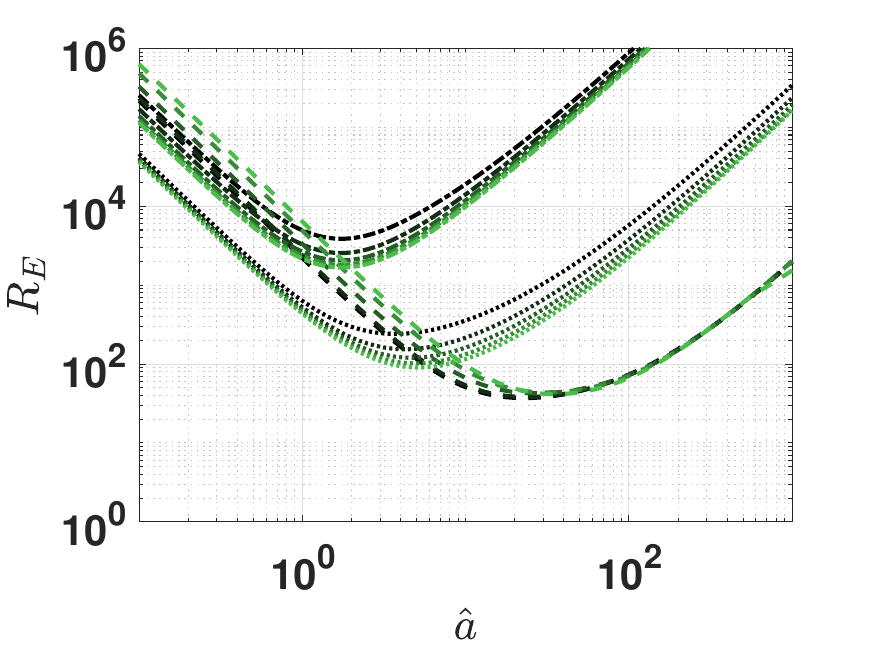}
	\includegraphics[width=0.49\textwidth]{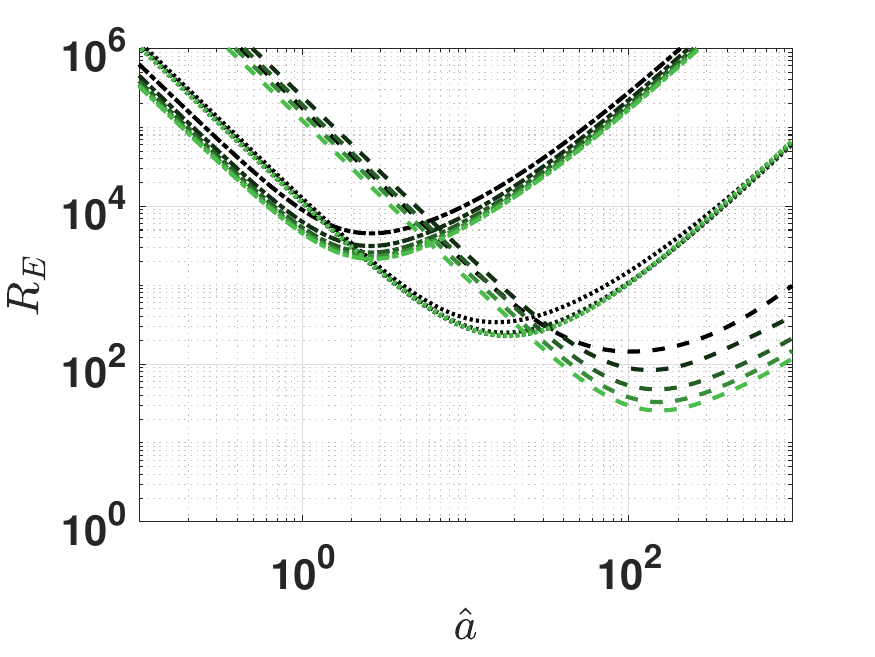}
	\caption{Rayleigh number $R_E$ as a function of angular wavenumber $\hat a$, using bottom pressure of $\hat P_B=1.25$ (left) and $\hat P_B=1.75$ (right), using constant viscosity (top row) or varying viscosity (bottom row). Lines corresponding to dashed, dotted and dash-dotted lines are for $\beta=0.01$, $\beta=0.1$ and $\beta=1$, respectively. Different colors are for times $\hat t=0.2,0.4,0.6,0.8$ and $\hat t=1.0$, where later times are in brighter green. }
	\label{fig:Rvsa}
\end{figure}

By solving the eigenvalue problem \eqref{eq:evp_simple} for various values of $\hat a$, $\hat t$ and $\beta$, we can find the behavior of the  Rayleigh number $R_E$ as a function of $\hat a$, $\hat t$ and $\beta$. Typical behaviors of $R_E$ with $\hat a$ for different $\hat t$ and $\beta$ are shown in Figure \ref{fig:Rvsa}. As observed in the figure, there is a value of $\hat a$ where the minimum value for $R_E$ occurs. This is then taken as $R_c$. In the case that we consider a horizontally bounded domain, \eqref{eq:arest} applies, which means that only certain values of $\hat a$ can be used. In this case, we only need to find $R_E$ for the allowed values of $\hat a$ and minimize over those, where the minimum is taken as $R_c$. Note that the critical Rayleigh number $R_c$ is hence expected to be larger when the domain is bounded. In both the bounded and unbounded case, the angular wavenumber that corresponds to the identified $R_c$, is taken as the critical angular wavenumber, and we can calculate a corresponding critical wavelength. In Figure \ref{fig:Rvsa} we have either considered the case of constant or varying viscosity. As the viscosity is increasing with salt concentration, the varying viscosity is expected to have a stabilizing effect on the instabilities. When using a constant viscosity, all terms including $\hat\mu'(\hat {\mathsf x}^0)$ in \eqref{eq:evp_simple} disappear.

\begin{figure}[h!]
	\centering
	\includegraphics[width=0.49\textwidth]{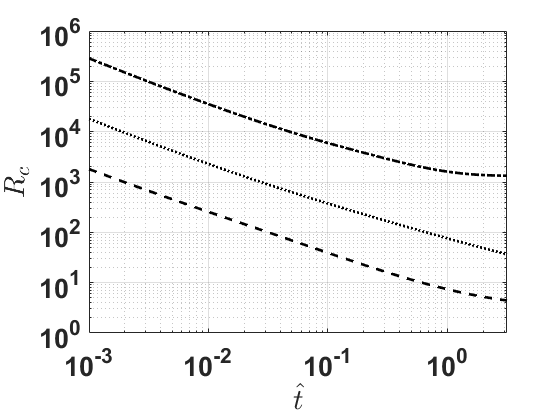}
	\includegraphics[width=0.49\textwidth]{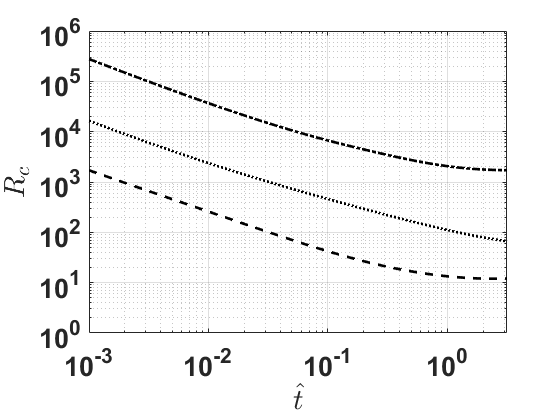}
	\includegraphics[width=0.49\textwidth]{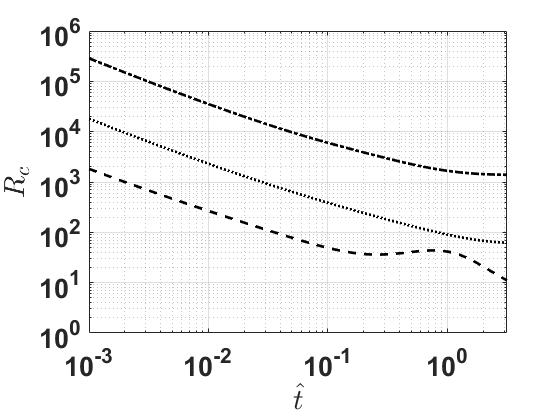}
	\includegraphics[width=0.49\textwidth]{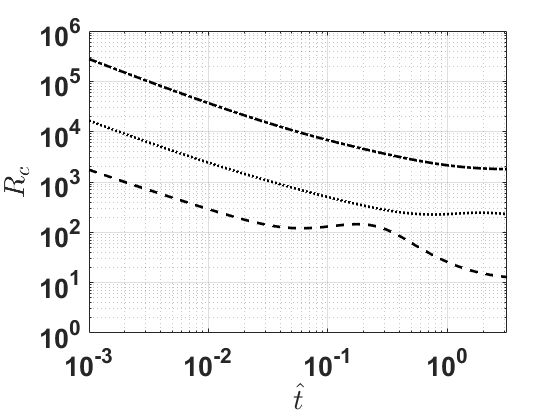}
	\caption{Critical Rayleigh number $R_c$ as function of time $\hat t$, using bottom pressure of $\hat P_B=1.25$ (left) and $\hat P_B=1.75$ (right), using constant viscosity (top row) or varying viscosity (bottom row). Lines corresponding to dashed, dotted and dash-dotted lines are for $\beta=0.01$, $\beta=0.1$ and $\beta=1$, respectively. }
	\label{fig:Rvst}
\end{figure}

From Figure \ref{fig:Rvsa} we also observe that for most cases the curves decrease with time. This means that the critical Rayleigh number decreases with time as well in these cases, which is shown in Figure \ref{fig:Rvst}. 
In practice, the system Rayleigh number $R_E^s = \frac{K\rho_\text{ref}g}{\mu_\text{ref}E}$ is a constant and pre-defined value. When the critical Rayleigh number $R_c(\hat t)$ decreases with time, there will be a critical time $\hat t_c$ where $R_c(\hat t)$ crosses this value. At this time, it is possible for non-trivial solutions to appear, meaning that instabilities will arise. This we call the onset time of the instabilities. 
When the viscosity is allowed to vary, there are two competing effects as the salt concentration increases: increased viscosity stabilizes while increased density destabilizes. For $\beta=0.01$ (corresponding to small diffusion), we can see in Figure \ref{fig:Rvst} that $R_c(\hat t)$ is not monotone in this case. 

In Figure \ref{fig:Rvst} we observe that for large values of $\beta$, the system is more stable as the critical Rayleigh numbers are larger. Correspondingly, a low value of $\beta$ results in a lower critical Rayleigh number. When the critical Rayleigh number decreases with time in a monotone way, we can find a unique onset time corresponding to a given Rayleigh number. When the critical Rayleigh number is not monotone with time, we would take the first time as where $R_c$ crosses the value of the given Rayleigh number as the onset time. Also note that the critical Rayleigh number also flattens out for some cases and does not appear to decrease any further with time, or decreases very slowly. This behavior corresponds to the ground state solution approaching its steady-state solution. Hence, there are systems where instabilities can never occur, when the criterion $R_E>R_c$ is not fulfilled.

Also note that for $\beta=0.01$, the appearing critical Rayleigh numbers are around 10 and even lower after around 0.5 time units for $\hat P_B=1.25$. As the eigenvalue problem relies on the simplified ground state which was found to not approximate the original ground state well for such low Rayleigh numbers, these results should not be trusted. However, for larger values of $\beta$, the corresponding critical Rayleigh numbers are in a range where the eigenvalue problem is a good approximation, at least in the time frames of interest.

\section{Numerical Model} \label{sec:numModel}

We use numerical simulations to review the results of the linear stability analysis and to determine the evolution of instabilities after the onset. For this, we use the  open-source simulator and research code \dumux (short for "DUNE for Multi-{Phase, Component, Scale, Physics, …} flow and transport in porous media") \cite{Koch2021} in which we have implemented for a two-dimensional bounded domain $\Omega$ the mathematical model described in Section \ref{sec:model}. 

The only difference in the initial conditions is that a small perturbation is added to the initial salt concentration. We add either a periodic or random distribution as perturbation to initiate the instabilities. Without any initial perturbation, the instabilities would be triggered by the numerical error \cite{BringedalSatInstab}. For the periodic perturbation, we add to the salt concentration a cosine function along the horizontal axis $x$: 

\begin{align} \label{eq:periodicPerturb}
	\mathsf{x}_{0,p}^\mathrm{NaCl}(x,z) = \mathsf{x}_0^\mathrm{NaCl} + A(z) \cdot \cos \left(\frac{2 \pi}{\lambda}\cdot x\right), 
\end{align}

with a given amplitude $A(z)$, which is depending on the depth $z$, and given wavelength $\lambda$. The wavelengths are chosen such that a fixed number of waves fit into the domain. For the random perturbation, we add to each grid node a salt concentration based on a normal distribution $\mathcal{N}$ defined by a mean $\overline{\mathsf{x}}^\mathrm{NaCl} = \mathsf{x}_0^\mathrm{NaCl}$ and a prescribed standard deviation $\sigma$:

\begin{align} \label{eq:randomPerturb}
	\mathsf{x}_{0,p}^\mathrm{NaCl} \thicksim \mathcal{N}\left(\overline{\mathsf{x}}^\mathrm{NaCl}, \sigma \right).
\end{align}

\subsection{Space and time discretization}

For the spatial discretization, we use a vertex-centered finite volume method, also known as the Box method, and for the time discretization an implicit Euler method \cite{Helmig2011}. Previous studies have suggested that the horizontal grid cell length, $\Delta x$, should be smaller than the expected wavelength \cite{Elenius2012, BringedalSatInstab}. As demonstrated by \cite{BringedalSatInstab} in a convergence study, a fine resolution in the vertical direction, $\Delta z$,  is essential as well to accurately capture the variation in salt concentration in the upper part of the domain. Therefore, we apply grid refinement in the vertical direction near the top of the domain.

As our simulation setup is similar to that presented in \cite{BringedalSatInstab}, we will use the same resolution of the spatial and temporal discretisation. For the spatial discretization, this means we have a horizontal grid cell length of $\Delta x = 0.001$ m and vertical grid cell heights that vary over the height of the domain starting with $\Delta z^{bottom} = 0.02 $ m  at the bottom of the domain and ending with $\Delta z^{top} = 3.33 \cdot 10 ^{-4}$ m at the top of the domain. For the temporal resolution, we choose a time-step size $\Delta t = 50$ s. This ensures the natural condition $\Delta z^{top} / \Delta t > E $. \\

\subsection{Influence of type of perturbation}\label{sec:numperturb}

We are interested in the onset of instabilities and on the number of waves $n_x$ at that moment. We define the onset as the time where the minimum of the standard deviation of the salt concentration in the top cell row $\sigma^{top}$ occurs. Additionally, we perform a fast Fourier transform to estimate which wavelength is appearing. We identify the number of waves present at onset as the dominant number of waves.

As in \cite{BringedalSatInstab,KlokerApproaches}, we will investigate the impact of applying different vertical profiles of perturbations. For the periodic case, this is done by selecting specific $A(z)$, while for the random case we either perturb the full domain or the top row. 
For this investigation, we use the setup of the base case from Section \ref{sec:results}.

\begin{figure}[h!]
	\centering
	\includegraphics[width=0.6\textwidth]{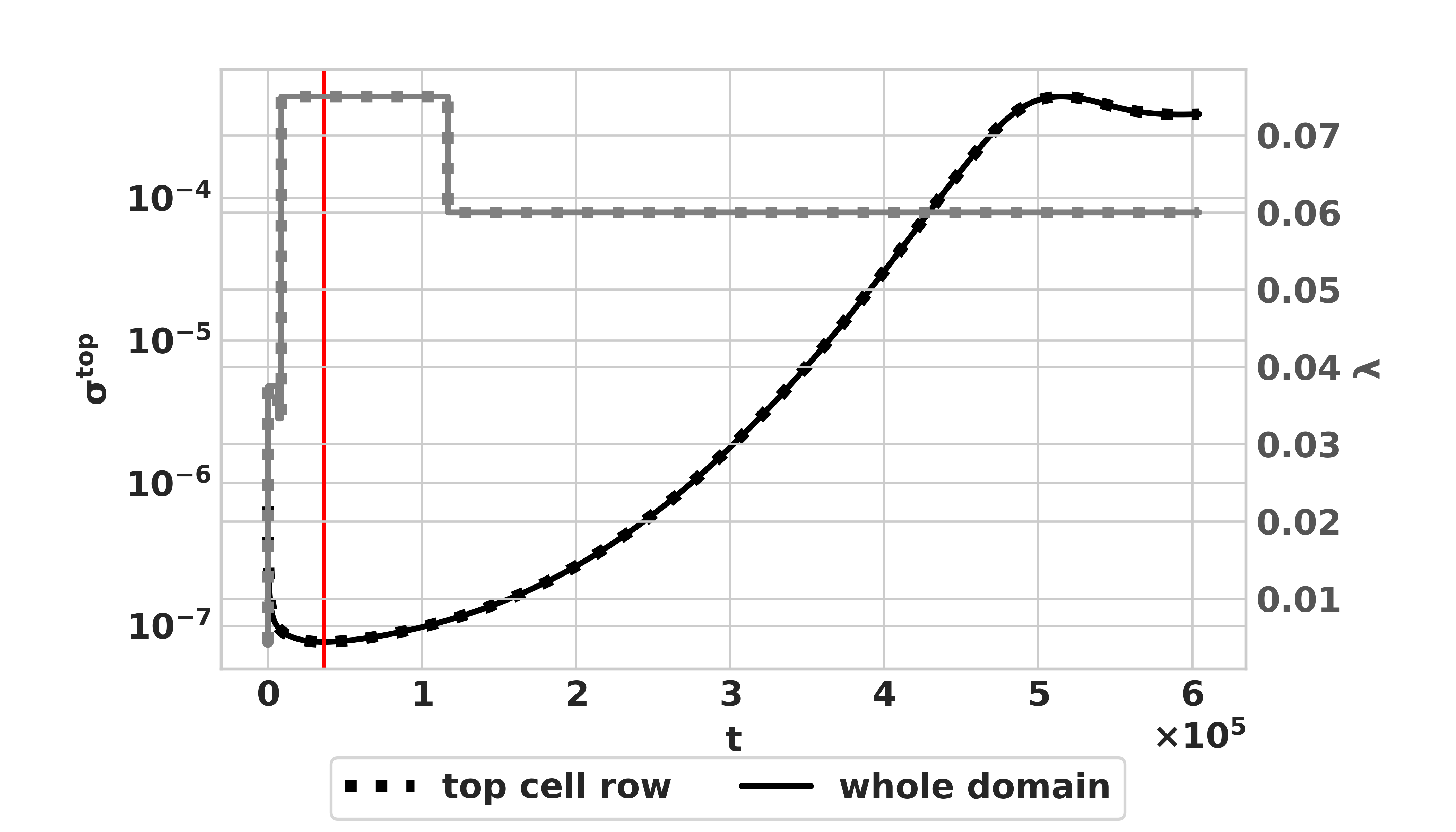}
	\caption{Influence of either perturbing the whole domain or the top row for the random case on the onset of instabilities (red line) which is derived by $\sigma^{top}$ (black line) and on the dominant wavelength $\lambda$ (grey line). }  
	\label{fig:Numerical_simulations_random_WL025}
\end{figure}

For the random perturbation, regardless of whether the domain is perturbed entirely or just in the top cell row, the results are the same for the base case. Figure \ref{fig:Numerical_simulations_random_WL025} shows that the results of both perturbation types yield the same $\sigma^{top}$ and the same wavelength $\lambda$, resulting in the same onset time and number of waves $n_x$.

\begin{figure}[h!]
	\centering
	\includegraphics[width=\textwidth]{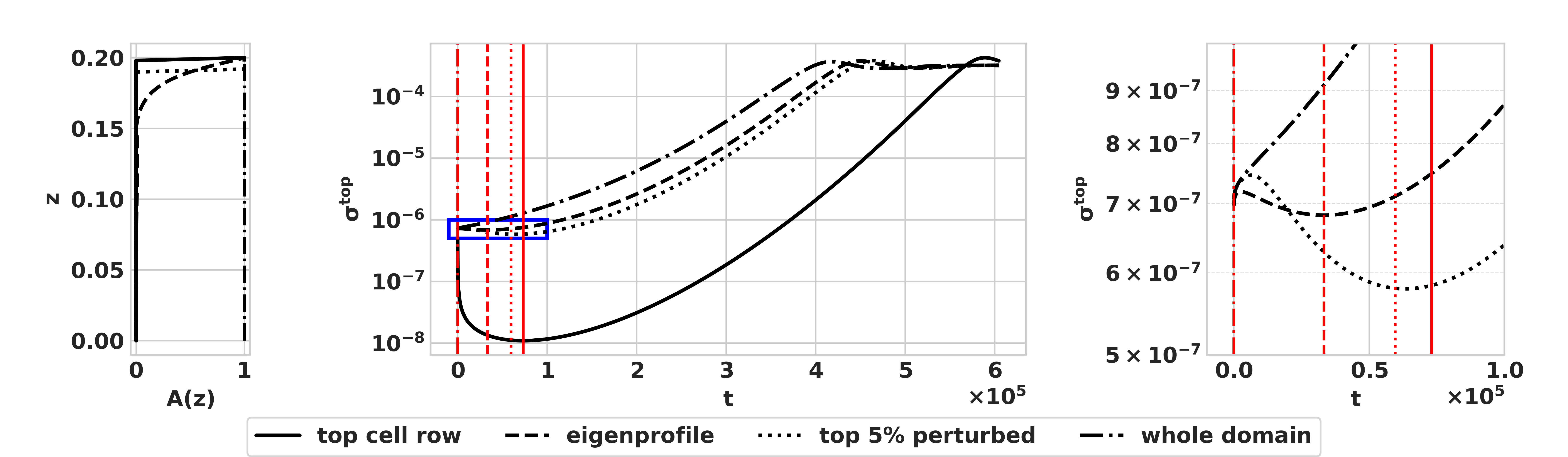}
	\caption{Impact of the region in which the periodic perturbation is applied on the numerical onset of instabilities. The red lines represent the respective onset times for each scenario.}
	\label{fig:Numerical_simulations_periodic_perturbationheight}
\end{figure}

From the random perturbation, we identify the dominant number of waves of the base case as $n_x$ = 8, which we apply as the initial number of waves for the periodic perturbation. We then analyze the influence of the vertical profile on the onset of instabilities by varying $A(z)$, see Figure \ref{fig:Numerical_simulations_periodic_perturbationheight}. We select four representative cases of different vertical profiles: (I) perturbing only the top cell row, (II) using the respective eigenprofile from the linear stability analysis, (III) perturbing the top 5$\%$, and (IV) perturbing the whole domain. 

When the entire domain is perturbed, the convective flux dominates the diffusive flux, resulting in a direct increase in the standard deviation. Thus, the onset of instabilities is at the start of the simulation. By changing the $A(z)$-profile so that only the upper domain is perturbed, the simulations show that the diffusive flux dominates at first. However, for the base case this is only if not more than 5$\%$ of the domain (III) is perturbed. The smaller the portion of the domain that is perturbed, the later the onset. Additionally, if choosing the vertical profile based on the eigenprofile from the linear stability analysis (II), the onset time occurs earlier than if we choose discontinuous $A(z)$-profiles. 

\section{Onset and Behavior of Density Instabilities}\label{sec:results}
The saturation influences the salt concentration profile, which is the driving force for the onset of instabilities.  
Therefore we reconsider the salt conservation equation \eqref{eq:NaCl}, where the saturation appears in three terms: (1) in the storage term, (2) in the convective term through the relative permeability, (3) in the diffusive term through the effective diffusion coefficient. In the following analysis we aim to isolate each of these three effects to investigate their importance for onset times separately, and afterwards consider their combined effect on the onset times. 

As a reference, we consider a base case with parameters as given in Table \ref{tab:values}. The hydraulic parameters of the porous medium, selected to represent fine sand, are obtained from \cite{Bechtold2011}. The corresponding dimensionless numbers and reference values are given in Table \ref{tab:nondimvalues}. This base case was investigated in Section \ref{sec:lsa_groundstate} and \ref{sec:lsa_sol} with $\hat P_B=1.25$. The chosen parameters yield a Rayleigh number for which the simplified eigenvalue problem \eqref{eq:evp_simple} is applicable. Note that we consider a horizontally bounded domain to allow comparison between the linear stability analysis and numerical simulations. This means that the restriction \eqref{eq:arest} on the angular wavenumber applies. 

\begin{table} [H]
	\begin{center}
		\begin{tabular}{lll}
			Symbol       & Value  & Dimension  \\
			\hline
			$\phi$ & 0.41 &  - \\
			$K$ & $2.89\cdot 10^{-11}$ & m$^2$ \\
			$D_\mathrm{w}^\mathrm{NaCl}$ & $1.5\cdot10^{-9}$ & m$^{2}$ s$^{-1}$\\ 
			$n$ & 10.8 & - \\
			$m$ & 0.907 & -\\
			$S_{wr}$ & 0.122& - \\
			$L$ & 0.73 & - \\
			$p_B$ & 2464 & kg m$^{-1}$ s$^{-2}$\\
			$\alpha$ & $1.77\cdot10^{-4}$ & (kg m$^{-1}$ s$^{-2}$)$^{-1}$\\
			$E$ & $1.08\cdot10^{-8}$ & m s$^{-1}$\\
			$H$ & 0.2 & m\\
			$\ell$ & 0.6 & m\\
			$g$ & 9.8 & m s$^{-2}$\\
			$\mathsf x_\mathrm{w}^\mathrm{NaCl}|_{t=0}$ & 0.0036 & mol mol$^{-1}$\\   
			\hline
		\end{tabular}
		\caption{Overview of values used in base case.}
		\label{tab:values}
	\end{center}
\end{table}

\begin{table} [H]
	\begin{center}
		\begin{tabular}{lll}
			Symbol       & Value  & Dimension \\
			\hline
			$p_\text{ref}$ & 1972 & Pa \\
			$Q_\text{ref}$ & $2.59\cdot 10^{-4}$ & m s$^{-1}$ \\
			$t_\text{ref}$ & $7.59\cdot 10^{6}$ & s  \\
			$R_E$ & 23 984 & - \\
			$\beta$ & 0.1823 & - \\ 
			$\gamma$ & 0.349 & - \\
			\hline
		\end{tabular}
		\caption{Overview of reference and non-dimensional values for base case.}
		\label{tab:nondimvalues}
	\end{center}
\end{table}

To address the influence of saturation, we propose the following: In Section \ref{sec:results_processes}, we consider the influence of varying storage, convection and diffusion separately. 
Then, in Section \ref{sec:results_unsat2}, the combined influence is investigated. In these two sections, we use the linear stability analysis. 
We compare the results from the linear stability analysis with results from the numerical simulations in Section \ref{sec:results_comp}. In Section \ref{sec:results_num}, the numerical simulations are used to investigate the further development of the instabilities.

\subsection{Varying storage, convection and diffusion}\label{sec:results_processes}

We want to understand the role of the water saturation on the onset time of instabilities For this we consider the salt transport equation \eqref{eq:NaCl} where the saturation appears in the storage, convection and diffusion term. We investigate changes in the relative strengths of these terms by adjusting other model parameters while keeping the saturation fixed. This way, we mimic the hypothetical influence a varying saturation has in the three individual terms, without considering the fully coupled system. This is done by altering the value of the porosity $\phi$ (in the storage term only), the intrinsic permeability $K$, and the diffusion coefficient $D_\mathrm{w}^\mathrm{NaCl}$, respectively. Each of these parameters are adjusted to three new values, which gives us in total 10 cases, when also considering the base case (Table \ref{tab:values}). The 9 cases with adjusted parameters are described in Table \ref{tab:changes}. 

\begin{table} [H]
	\begin{center}
		\begin{tabular}{llll}
			Case Number & Parameter   & New value  & Dimension  \\
			\hline
			Case 1 & $\phi$ & 0.37 &  - \\
			Case 2 & $\phi$ & 0.30&  - \\
			Case 3 & $\phi$ & 0.21 &  - \\ \hline
			Case 4 & $K$ & $1.94\cdot 10^{-11}$ & m$^2$ \\
			Case 5 & $K$ & $8.62\cdot 10^{-12}$ & m$^2$ \\
			Case 6 & $K$ & $2.35\cdot 10^{-12}$ & m$^2$ \\ \hline
			Case 7 & $D_\mathrm{w}^\mathrm{NaCl}$ & $1.0\cdot10^{-9}$ & m$^{2}$ s$^{-1}$\\ 
			Case 8 & $D_\mathrm{w}^\mathrm{NaCl}$ & $4.8\cdot10^{-10}$ & m$^{2}$ s$^{-1}$\\ 
			Case 9 & $D_\mathrm{w}^\mathrm{NaCl}$ & $1.5\cdot10^{-10}$ & m$^{2}$ s$^{-1}$\\ 
			\hline
		\end{tabular}
		\caption{Overview of values that are changed compared to the base case in the 9 other cases.}
		\label{tab:changes}
	\end{center}
\end{table}

The values in Table \ref{tab:changes} have been chosen in the following way: By increasing the bottom pressure $P_B$, a new (initial) saturation results, with a corresponding saturation at the top of the domain. This new top saturation, compared to the top saturation of the base case, has been used to calculate a corresponding change in apparent porosity, intrinsic permeability and intrinsic diffusion, by the expected change in saturation, relative permeability and effective diffusion, respectively. Case 1, 4 and 7 correspond to a change in bottom (capillary) pressure from 2464 Pa to 2958 Pa, Case 2, 5 and 8 to 3451 Pa, and Case 3, 6 and 9 to 3944 Pa. These changes in bottom pressure lead to a gradual decrease in saturation and therefore reduction in apparent porosity, intrinsice permeability and intrinsic diffusion. Note that the porosity change for Case 1, 2 and 3 are only done for the storage term in \eqref{eq:NaCl}. However, as already pointed out: the hypothetical change in saturation is only used to calculate new values of porosity, permeability and intrinsic diffusion. This way, Case 1, 2 and 3 mimic the influence of changes in the storage term, Case 4, 5 and 6 mimic changes in the convective term, and Case 7, 8 and 9 mimic changes in the diffusive term. 

The changes in parameters translate to a change in the reference time $t_\text{ref}$ (when changing $\phi$), $R_E$ (when changing $K$) and $\beta$ (when changing $D_\mathrm{w}^\mathrm{NaCl}$). The values of $t_\text{ref}, R_E$ and $\beta$ for the base case are given in Table \ref{tab:nondimvalues} and changes made for Case 1-9 are given in Table \ref{tab:nondimchanges}. 

\begin{table} [H]
	\begin{center}
		\begin{tabular}{llll}
			Case Number & Parameter   & New value  & Dimension  \\
			\hline
			Case 1 & $t_\text{ref}$ & $6.83\cdot10^6$ &  s \\
			Case 2 & $t_\text{ref}$ & $5.48\cdot10^6$ &  s \\
			Case 3 & $t_\text{ref}$ & $3.91\cdot10^6$ &  s \\ \hline
			Case 4 & $R_E$ & 16 108 & - \\
			Case 5 & $R_E$ & 7154 & -\\
			Case 6 & $R_E$ & 1950 & - \\ \hline
			Case 7 & $\beta$ & 0.1252 & -\\ 
			Case 8 & $\beta$ & 0.0581 & -\\ 
			Case 9 & $\beta$ & 0.0177 & -\\ 
			\hline
		\end{tabular}
		\caption{Overview of values that are changed compared to the base case in the 9 other cases.}
		\label{tab:nondimchanges}
	\end{center}
\end{table}

\subsubsection{Fully saturated domain}\label{sec:results_sat}
We first consider a fully saturated domain ($S_\mathrm{w}=1$). This allows us to lean on previous works: see \cite{BringedalSatInstab,KlokerApproaches} and \cite{BringedalEigenvalue,KlokerCode} for the codes of the eigenvalue problems. Note that in \cite{BringedalSatInstab,KlokerApproaches} (and in the codes), the viscosity was taken constant. Hence, the influence of varying viscosity is not accounted for here.

\begin{table} [H]
	\begin{center}
		\begin{tabular}{l|cc|cc}
			\multirow{2}{*}{Case number} & \multicolumn{2}{c|}{Unbounded}   & \multicolumn{2}{c}{Bounded} \\
			& Onset time & $n_x$ & Onset time & $n_x$ \\
			\hline
			Base & 10 830 s & 1 & 10 791 s & 1\\ \hline
			Case 1 & 9747 s & 1 &  9712 s & 1\\
			Case 2 & 7819 s & 1 &  7790 s & 1\\
			Case 3 &  5574 s & 1 &  5553 s & 1 \\ \hline
			Case 4 & 15 709 s & 1 & 16 411 s & 1\\
			Case 5 & 37 790 s & 1 & 39 165 s & 1\\
			Case 6 & 160 525 s & 1 & 167 359 s & 1\\ \hline
			Case 7 & 6933 s & 1 & 6680 s & 1\\ 
			Case 8 & 2922 s & 1 & 3109 s & 1\\ 
			Case 9 & 834 s & 1 & NaN & NaN\\ 
			\hline
		\end{tabular}
		\caption{Onset times and preferred number of waves $n_x$ for fully saturated vertically unbounded and bounded domains.}
		\label{tab:satresults}
	\end{center}
\end{table}

Table \ref{tab:satresults} contains the onset times and corresponding number of waves (which is directly linked to the critical wavelength and critical angular wavenumber through the domain width) for the base case and Case 1-9. Note that Case 9 for the bounded setup caused numerical instabilities in the eigenvalue problem solver, causing this case to not provide any results. However, from the other results we can observe clear trends: Lowering the apparent porosity in the storage term (Case 1-3) gives slightly earlier onset times. Hence, lower storage capacity has a slight destabilizing effect on the system as the density instabilities can appear earlier. Lowering the intrinsic permeability (Case 4-6) gives much later onset times. The lower strength of the convective term has a strong stabilizing effect on the system. Finally, lowering the intrinsic diffusion (Case 7-9) results in much earlier onset times. This means that lower diffusion has a strong destabilizing effect on the system. For all these cases, the preferred number of waves is always one. For the vertically unbounded domain, this is always the preferred number (see discussion in \cite{BringedalSatInstab}), while the preferred number is calculated for each case for the vertically bounded domain.

\subsubsection{Partially saturated domain: fixed saturation}\label{sec:results_unsat}
Next we consider a partially saturated domain with a fixed (i.e.~time-independent saturation). 
The actual saturation profile will be the same in all the cases and is the base case and can be seen in red in the top right of Figure \ref{fig:all_GS_pbot125}. The eigenvalue problem \eqref{eq:evp_simple} is solved, and we modify the corresponding value of $t_\text{ref}$ (when changing $\phi$), $R_E$ (when changing $K$) and $\beta$ (when changing $D_\mathrm{w}^\mathrm{NaCl}$) for the Case 1-9. The corresponding dimensional onset times and preferred number of waves are given in Table \ref{tab:unsatresults}. Here we allow for either constant or varying viscosity, to address also the influence the varying viscosity has on the onset times.

\begin{table} [H]
	\begin{center}
		\begin{tabular}{l|cc|cc}
			\multirow{2}{*}{Case number} & \multicolumn{2}{c|}{Constant viscosity}   & \multicolumn{2}{c}{Varying viscosity} \\
			& Onset time & $n_x$ & Onset time & $n_x$ \\
			\hline
			Base & 11 654 s & 1 & 11 671 s & 1\\ \hline
			Case 1 & 10 489 s & 1 & 10 504 s & 1 \\
			Case 2 & 8414 s & 1 & 8426 s & 1 \\
			Case 3 & 5998 s & 1 & 6007 s & 1\\ \hline
			Case 4 & 17 804 s & 1 & 17 850 s & 1 \\
			Case 5 & 43 081 s & 1 & 43 294 s & 1 \\
			Case 6 & 193 211 s & 1  & 195 669 s & 1\\ \hline
			Case 7 & 7317 s & 1 & 7332 s & 1 \\ 
			Case 8 & 3098 s & 1 & 3105 s & 1\\ 
			Case 9 & 856 s & 3 & 858 s & 3 \\ 
			\hline
		\end{tabular}
		\caption{Onset times and preferred number of waves $n_x$ for partially saturated porous domain.}
		\label{tab:unsatresults}
	\end{center}
\end{table}

Note that the choices of Rayleigh number $R_E$ in Table \ref{tab:nondimvalues} and \ref{tab:nondimchanges} are all well above 10. Since only Rayleigh numbers from order of magnitude $10^3$ and larger are considered, the simplified eigenvalue problem \eqref{eq:evp_simple} applies and gives trustworthy results. 

From the onset times we see a similar trend as in Section \ref{sec:results_sat}. Compared to the base case, a decrease of the storage capacity (Case 1-3) has a slight destabilizing effect and gives somewhat earlier onset times. Decreasing the strength of the convective term (Case 4-6) has a strong stabilizing effect, while decreasing the strength of the diffusion (Case 7-9) has a strong destabilizing effect. Note that lowering the intrinsic permeability has an even stronger effect for the partially saturated domain compared to the fully saturated. The influence of lowering the storage and diffusion appears to be comparable between the partially and fully saturated domains. The non-dimensional onset times are all below 0.026 (maximum found for Case 6), meaning that the changes observed in Section \ref{sec:lsa_groundstate} are more extreme than considered here. The influence of varying viscosity is minimal for the parameter choices considered here. The onset times are later when the viscosity is varying with salt concentration, confirming that a higher viscosity has a stabilizing influence, but the differences in onset times are marginal. The preferred number of waves at onset is not influenced by the viscosity.

\subsection{Partially saturated domain: varying saturation}\label{sec:results_unsat2}
\begin{figure}[H]
	\centering
	\includegraphics[width=0.49\textwidth]{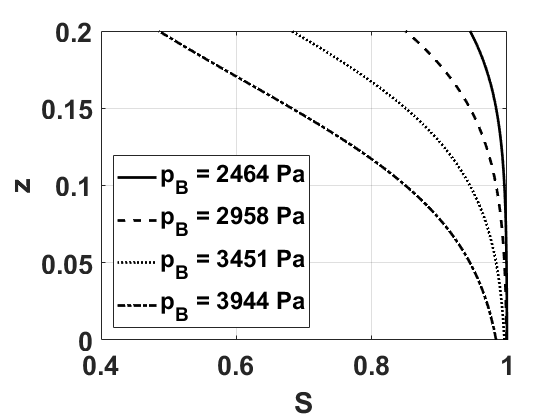}
	\caption{Saturation profiles for the four different bottom pressures, corresponding to four different bottom pressures.}
	\label{fig:Sprofiles}
\end{figure}

We now combine all three effects and consider different saturation profiles in the ground state. We keep all parameters as specified in Table \ref{tab:values} and \ref{tab:nondimvalues}, but vary the bottom pressure. The base case uses $p_B = 2464$ Pa, and we consider additionally $p_B=2958$ Pa, $p_B=3451$ Pa and $p_B=3944$ Pa. The different saturation profiles are shown in Figure \ref{fig:Sprofiles}. Now, the three effects will compete: as the saturation is lower, storage, convection and effective diffusion will be lower. But where lower storage and diffusivity destabilize the system and would result in an earlier onset time, lower convection does the opposite. The resulting onset times for these four cases are given in Table \ref{tab:unsatresults2}.

\begin{table} [H]
	\begin{center}
		\begin{tabular}{l|cc|cc}
			\multirow{2}{*}{Bottom pressure}  & \multicolumn{2}{c|}{Constant viscosity}   & \multicolumn{2}{c}{Varying viscosity} \\
			& Onset time & $n_x$ & Onset time & $n_x$ \\
			\hline
			2464 Pa (base) & 11 654 s & 1 & 11 671 s & 1 \\ \hline
			2958 Pa &  12 415 s & 1 & 12 452 s & 1 \\
			3451 Pa &  11 053 s & 3 & 11 110 s & 3\\
			3944 Pa & 7951 s & 6 & 8042 s & 6  \\ \hline
		\end{tabular}
		\caption{Onset times and number of waves for partially saturated porous domain.}
		\label{tab:unsatresults2}
	\end{center}
\end{table}

As observed in Table \ref{tab:unsatresults2}, there is no clear trend when the saturation is lowered. The onset times first become a bit later, and then earlier. Hence, the combined influence of the three effects appears to be that lowering convection dominates for a small decrease in saturation, while lowering storage and diffusivity dominates as the saturation decreases further. Notably, a larger number of waves (corresponding to a shorter wavelength) is found to be preferred for the two lowest saturations, although an increased number of waves was only found for Case 9 (lowest diffusion) in the previous analysis. It is possible that the combined influence of the three effects has led to even increased number of waves being preferred. Another possibility is that having a stronger vertical profile itself leads to a preference towards increased number of waves. Finally, the influence of varying viscosity is minimal and has only a marginal stabilizing effect on the onset of instabilities. 

\subsection{Comparison of onset times between linear stability analysis and numerical simulations}\label{sec:results_comp}

To compare the onset times from the linear stability analysis with the onset times found via numerical simulations of the full model, we take a two-step approach. 
In the first step we randomly perturb the constant initial salt concentration with $\sigma = 10^{-6} $mol mol$^{-1}$, and numerically determine the resulting concentration and discharge field. We do this for each of the four bottom pressures. From the evolving salt profiles we determine the dominant number of waves and corresponding wavelength. We also determine for each case the onset time for the growth of instabilities. Here the onset time is defined as the time at which the standard deviation at the top of the domain ($z=H$)

\begin{equation}\label{eq:stdev}
	\sigma^\text{top}(t) = \sqrt{\int_0^\ell (\mathsf{x}_\mathrm{w}^\text{NaCl}(x,H,t)-\overline{\mathsf{x}}^\text{top}(t))^2dx}
\end{equation}

attains its minimum. We take here $z=H$ because here we expect the largest gradients in the salt concentraion. In \eqref{eq:stdev} we use the mean concentration

\begin{equation}
	\overline{\mathsf{x}}^\text{top}(t) = \frac{1}{\ell}\int_0^\ell \mathsf{x}^\text{NaCl}_\mathrm{w}(x,H,t)dx.
\end{equation}

The results are in Table \ref{tab:unsatresults_random}.

\begin{table} [H]
	\begin{center}
		\begin{tabular}{l|cc}
			{Bottom pressure}  & Onset time  & $n_x$  \\
			\hline
			2464 Pa (base) &   36 400 s &  8  \\ \hline
			2958 Pa &  30 350 s &  8  \\
			3451 Pa &  14 550 s &  18 \\
			3944 Pa & 10 100 s & 16  \\  \hline
		\end{tabular}
		\caption{Onset times and dominant number of waves when using a random perturbation for the numerical simulations}
		\label{tab:unsatresults_random}
	\end{center}
\end{table}

\begin{figure}[H]
	\centering
	\includegraphics[width=0.49\textwidth]{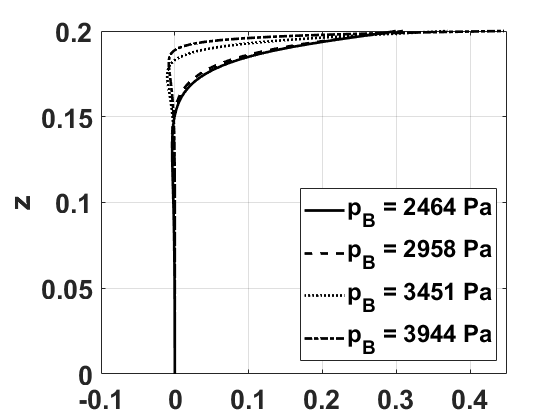}
	\caption{Eigenprofiles for the four different bottom pressures, corresponding to four different number of waves found to be dominant for the numerical simulations.}
	\label{fig:eigenprofiles}
\end{figure}

In the second step, we use the number of dominant waves, $n_x$, to construct horizontal periodic perturbations for which the onset times are compared by both simulations and linear stability analysis. The approach is motivated by the procedure in \cite{BringedalSatInstab}. As the amplitude $A(z)$ of the perturbation effects the onset times, see Section \ref{sec:numperturb} and \cite{KlokerApproaches}, we apply either a perturbation in the top row cells, or use for $A(z)$ the vertical eigenprofile, see Figure \ref{fig:eigenprofiles}, obtained by solving the eigenvalue problem \ref{eq:evp_simple}. We carry out the second step for each of the four bottom pressures. The results are given in Table \ref{tab:unsatresults_wave}.

\begin{table} [H]
	\begin{center}
		\begin{tabular}{ll|cc|c}
			\multirow{2}{*}{Bottom pressure} & \multirow{2}{*}{$n_x$} & \multicolumn{2}{c|}{Numerical simulation}   & {Linear stability} \\
			& & Top row & Eigenprofile  & analysis  \\
			\hline
			2464 Pa (base) &  8 &  73 050 s & 33 250 s & 22 693 s \\ \hline
			2958 Pa &  8 &  65 400 s & 26 150  s & 19 317 s \\
			3451 Pa &  18 &  38 500 s & 23 650 s & 20 899 s\\
			3944 Pa &  16 & 27 400 s & 6 750 s & 10 100 s  \\  \hline
		\end{tabular}
		\caption{Onset times for numerical simulations and linear stability analysis considering a specific number of waves. The numerical simulation considers two different types of perturbation; either perturbing only at the top grid cells, or using the eigenprofile from Figure \ref{fig:eigenprofiles}.}
		\label{tab:unsatresults_wave}
	\end{center}
\end{table}

As seen in Table \ref{tab:unsatresults_wave}, when using a similar perturbation (periodic with a comparable number of waves), the numerical simulations and linear stability predict somewhat similar onset times for the instabilities as the bottom pressure and saturation profile changes. In particular, when using the corresponding eigenprofile for the vertical structure of the perturbation, the onset times compare well to the ones found from the linear stability analysis. The overall trend is that the onset times are earlier when the saturation is lower.

\subsection{Development of instabilities}\label{sec:results_num}

\begin{figure}[h!]
	\centering
	\includegraphics[width=0.9\textwidth]{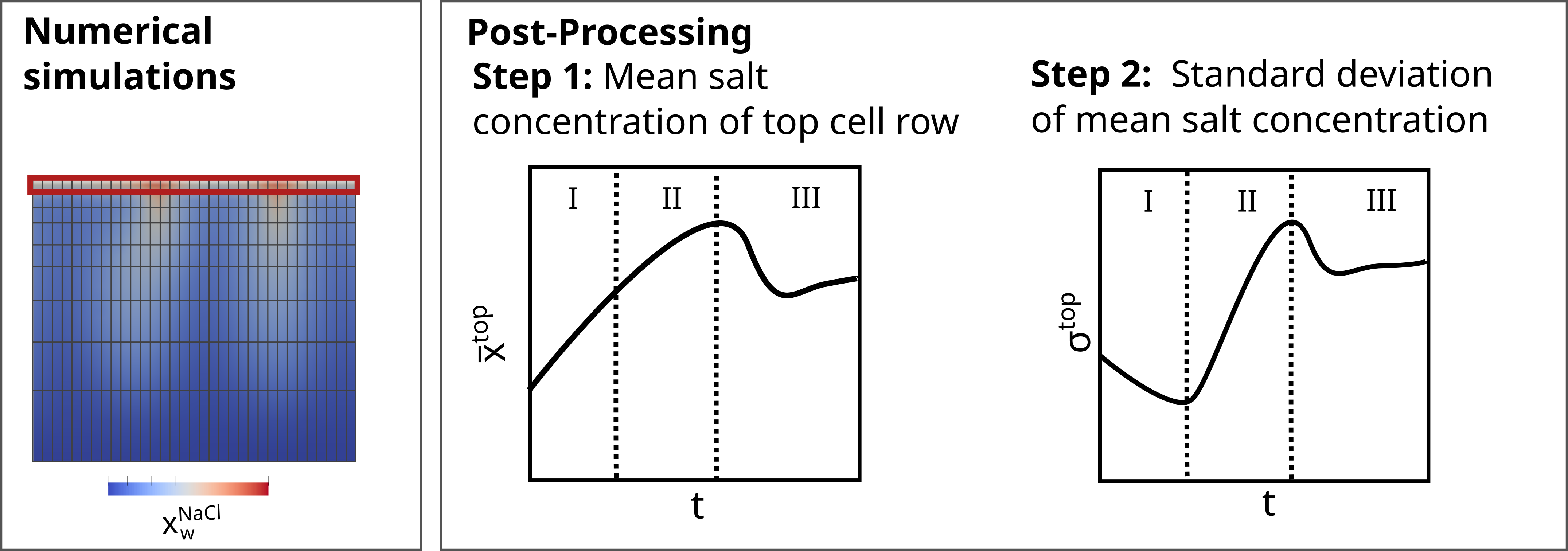}
	\caption{Schematic representation of the evaluation process for onset times from numerical simulations. a) Example output from the numerical simulation, with only the top row of the spatial discretization used for analysis. b) Workflow for onset time evaluation: Step 1 - Calculation of the mean concentration $\overline{\mathsf{x}}^\mathrm{top}$ at each time step; Step 2 - Determination of the standard deviation $\sigma^{top}$at each time step. The onset time is identified as the transition point between phase I and phase II. }
	\label{fig:Numerical_simulations_postprocessing_scheme}
\end{figure}

To interpret the behaviour of the salt concentration during the evolution of the evaporation process, we distinguish three phases. They are expressed in terms of $\overline{\mathsf{x}}^\mathrm{top}$ and $\sigma^{top}$, and schematically presented in Figure \ref{fig:Numerical_simulations_postprocessing_scheme}. This is similarly described in \cite{Slim_2014}.

\begin{itemize}
	\item Phase I: Initially, we start with a small perturbation to initiate instabilities. If diffusion dominates convection, the standard deviation decreases at first, while the salt concentration increases due to induced evaporation at the top. \\
	
	\item Phase II: As evaporation progresses, salt accumulates at the top of the domain. At a certain point, the convection takes over, causing fluctuations in the salt concentration to grow and the standard deviation to increase.\\
	
	\item Phase III: The salt concentration growth at the top will continue until the flow becomes gravitationally unstable, triggering a downward flow. This instability causes the mean salt concentration to decrease temporarily. As finger-like patterns develop in the flow, the mean salt concentration will oscillate, alternating between increases and decreases over time.\\
\end{itemize}

\begin{figure}[h!]
	\centering
	\includegraphics[width=\textwidth]{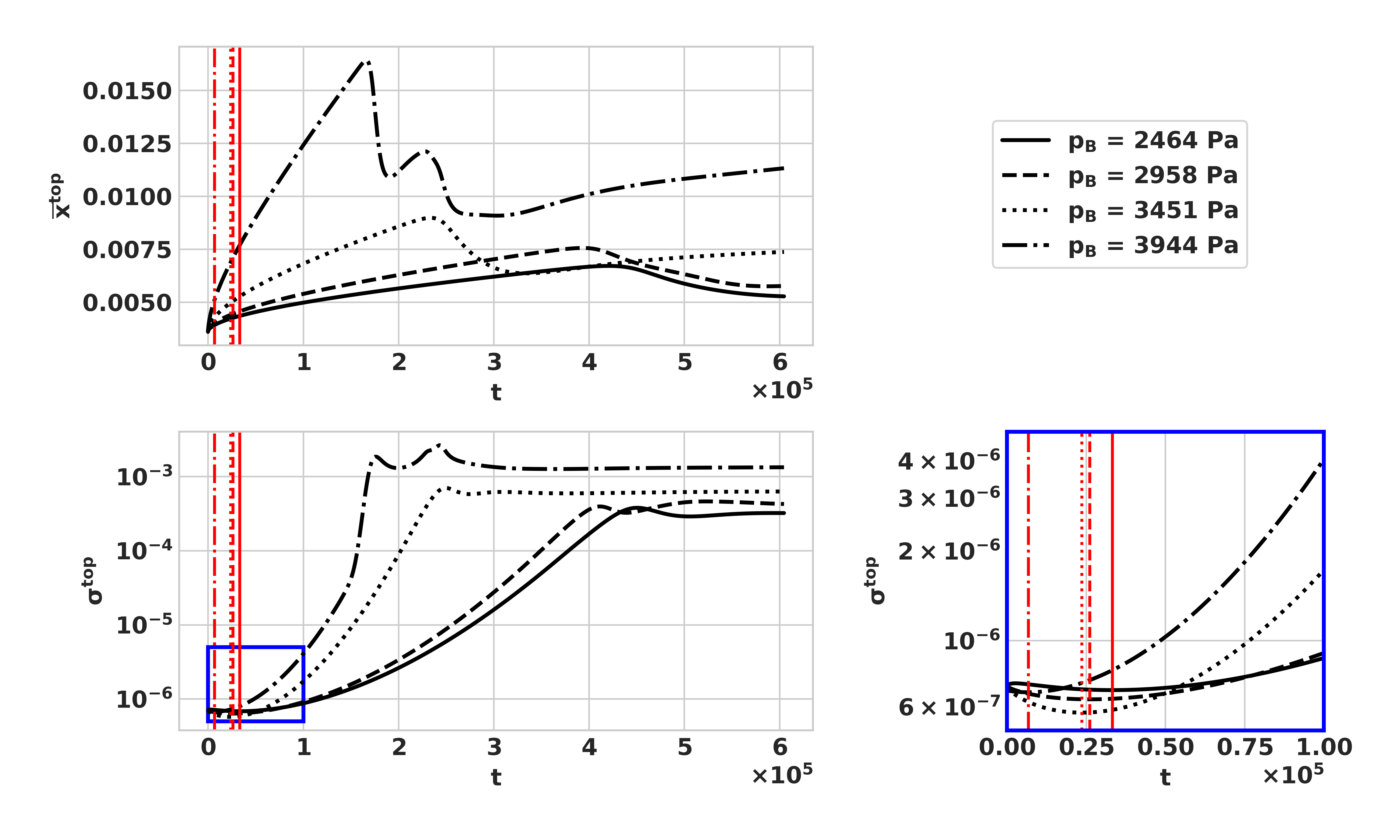}
	\caption{Further development of density instabilities using the periodic perturbation along the eigenprofile for different bottom pressures (${p_B}$). The red lines represent the corresponding onset times from table \ref{tab:unsatresults_wave}. }
	\label{fig:Numerical_simulations_periodic_eigenprofile_pertrubation}
\end{figure}

\begin{figure}[h!]
	\begin{subfigure}{\textwidth}
		\centering
		\includegraphics[width=0.7\textwidth]{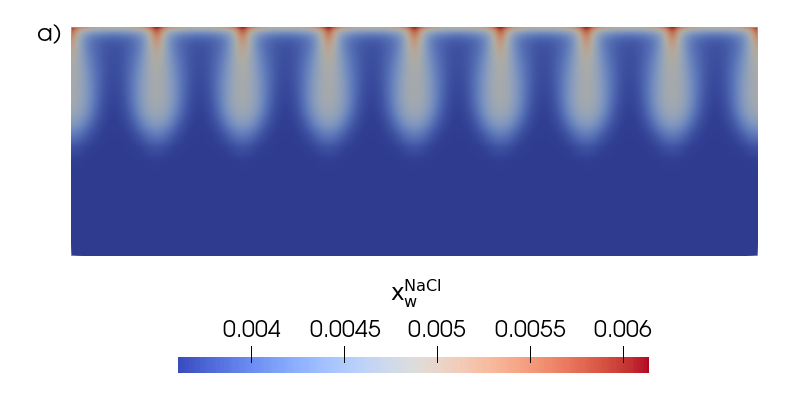}
	\end{subfigure}
	\begin{subfigure}{\textwidth}
		\centering
		\includegraphics[width=0.7\textwidth]{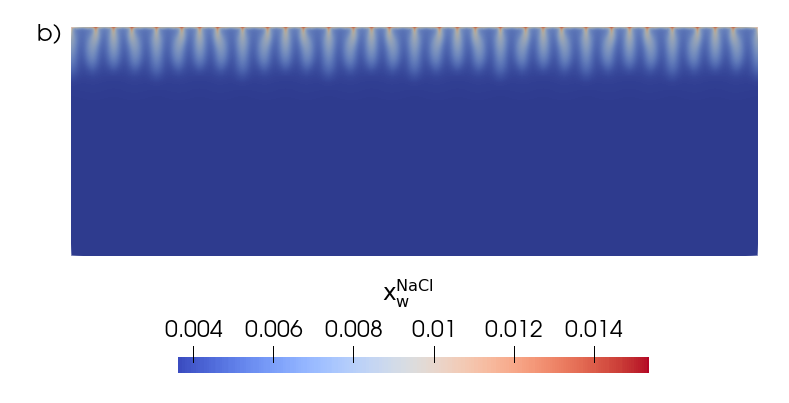}
	\end{subfigure}
	\caption{Finger pattern of the density instabilities at $t=6\times10^5$ s when using a) ${p_B} $ = 2464 Pa and b) ${p_B}$ = 3944 Pa. The concentration ranges of individual figures have been adjusted for improved visibility. }
	\label{fig:Simulation_DensityFingers}
\end{figure}

For the comparison with the linear stability analysis, we are interested in the transition from phase I to II. Further, by considering the numerical simulations in phase II and III, we gain insight in the progression of instabilities after the onset time. In particular we observe the formation of fingers for the four different bottom pressures. 
The initiation of the downward flow, marking the onset of finger development and the transition into phase III, is indicated by the peak in the mean salt concentration, $\mathsf{\overline{x}}^\mathrm{top}$. This behavior occurs consistently across all four cases (${p_B}$ = 2464 Pa to ${p_B}$ =3944 Pa), as illustrated in Figure \ref{fig:Numerical_simulations_periodic_eigenprofile_pertrubation}.  Similarly to the onset of instabilities, the onset of the downward flow occurs earlier for lower saturations. The lower saturation leads here also to a higher and also faster enrichment of the salt concentration at the top. Corresponding to Figure \ref{fig:Numerical_simulations_periodic_eigenprofile_pertrubation}, in Figure \ref{fig:Simulation_DensityFingers} the fingering due to downward flow using ${p_B}$ = 2464 Pa and ${p_B}$ = 3944 Pa at the end of the simulation is shown. The fingering pattern remains localized to the upper part of the domain for lower saturations. However, at higher saturations, the fingers extend deeper, spreading into the lower levels of the domain. Additionally, the onset of downward flow occurs earlier at lower saturations, leading to an observable merging of the fingers at this stage. In contrast, for increased saturations, the downward flow develops later, and the fingers retain their initial patterned structure without significant merging.

\section{Conclusion and Outlook}\label{sec:final}
In this work we have considered evaporation from a partially saturated porous medium, where a salt is dissolved in the water phase. As water evaporates from the top boundary of the porous medium, the salt accumulates near the top boundary giving possibly a gravitational unstable setting. By performing a linear stability analysis and numerical simulations of the full model, we have investigated the onset of instabilities, and via numerical simulations we could also address the progression of these instabilities.

The linear stability analysis requires further restrictions on the model equations. To be able to solve the resulting eigenvalue problem, we assumed that the saturation remained fixed with time. Numerical simulations of the full model equations, assured us that the saturation changes were very small in the parameter range and time period of interest. The resulting eigenvalue problem is very fast to solve, and can provide onset times for a wide range of parameter choices, as illustrated in Figure \ref{fig:Rvst}. The eigenvalue problem enables us to quickly identify interesting parameter choices that can be investigated further with numerical simulations of the full system. By solving the eigenvalue problem, we can determine when onset of instabilities occur, and in particular, whether onset of instabilities occur before salt precipitation starts.

Compared to the fully saturated case, the onset of instabilities in a partially saturated case is much more coupled as the saturation influences the storage, convection and effective diffusion of the salt. The linear stability analysis helped us to identify the influence of these processes and showed that reduction in storage has a small destabilizing influence (i.e., slightly earlier onset times), reduction in convection has a strong stabilizing influence, while reduction in effective diffusion has a strong destabilizing influence. However, when saturated is lower, all these influences will compete.

The linear stability analysis could show that the influence of varying viscosity is negligible for the onset of instabilities. Furthermore, the analysis revealed that the competing effects between reduction in storage, convection and diffusion are rather similar when the saturation reduces slightly, as the onset times are rather similar when saturation changes moderately. However, for stronger reduction in saturation, the onset times are much earlier, which we can connect to the reduction in diffusion and to some extent in storage. The onset times predicted by the numerical simulations are influenced by how the initial condition is perturbed. The numerical simulations still give a clear trend in earlier onset times as the saturation decreases, independent of the type of perturbation used. When using an eigenprofile perturbation in the numerical simulations, we obtain similar onset times as in the linear stability analysis. This shows that although these two approaches are based on different approaches and assumptions, they can give comparable onset times.

The numerical simulations revealed information on the further development of the instabilities after their onset. After onset of instabilities, the salt concentration at the top of the porous domain still increases as the instabilities are initially weak in strength. When the instabilities are sufficiently strong, we observe the formation of downwards propagating fingers that give a reduction in the salt concentration at the top. The formation of these fingers also depends on saturation. A lower saturation results in an earlier formation of fingers, where the fingers are generally small but merge gradually as time progresses. For higher saturation, the fingers form at a later point but extend deeper into the domain.

The analysis presented in this work shows the influence of varying saturation for evaporation-induced density instabilities. Although we considered one type of sand and therefore a specific choice of van Genuchten parameters, the analysis in general and can handle other types of relations as well as other properties for the liquid and dissolved salt. We did not include salt precipitation in the analysis, and all considered examples had properties such that salt would not precipitate in the time frames we considered. However, as the salt concentration is still increasing after the onset of instabilities, since the instabilities are initially too weak in strength to give a net downwards transport of salt, investigating the interplay between density instabilities and salt precipitation is a natural extension of the current work.

\section*{Statements and Declarations}

\begin{itemize}
\item Funding\\
We thank the Deutsche Forschungsgemeinschaft (DFG, German Research Foundation) for supporting this work by funding SFB 1313, Project Number 327154368.\\

\item Conflict of interest/Competing interests\\
We declare no conflict of interest. \\

\item Code availability \\
The code for the linear stability analysis (doi:10.18419/darus-4711) and for the numerical simulations (doi:10.18419/darus-4610) can be found in the data repository of the University of Stuttgart (DaRUS) \cite{Darus_2020}. 
\item Author contribution \\
All authors contributed to the study conception and design. Material preparation, data collection and analysis were performed by Carina Bringedal, Stefanie Kiemle, C.J. van Duijn and Rainer Helmig. The first draft of the manuscript was written by Carina Bringedal and Stefanie Kiemle, and all authors commented on previous versions of the manuscript. All authors read and approved the final manuscript.

\end{itemize}

\pdfbookmark[0]{Bibliography}{Bibliography}

\bibliographystyle{abbrvurl}

\bibliography{literature}

\begin{thebibliography}{10}

\bibitem{allison1985}
G.~Allison and C.~Barnes.
\newblock Estimation of evaporation from the normally “dry” lake frome in
  south australia.
\newblock {\em Journal of Hydrology}, 78(3):229--242, 1985.
\newblock \href {https://doi.org/10.1016/0022-1694(85)90103-9}
  {\path{doi:10.1016/0022-1694(85)90103-9}}.

\bibitem{barletta2009robin}
A.~Barletta, M.~Celli, and D.~A.~S. Rees.
\newblock The onset of convection in a porous layer induced by viscous
  dissipation: A linear stability analysis.
\newblock {\em International Journal of Heat and Mass Transfer},
  52(1):337--344, 2009.
\newblock \href {https://doi.org/10.1016/j.ijheatmasstransfer.2008.06.001}
  {\path{doi:10.1016/j.ijheatmasstransfer.2008.06.001}}.

\bibitem{Batzle1992}
M.~Batzle and Z.~Wang.
\newblock Seismic properties of pore fluids.
\newblock {\em GEOPHYSICS}, 57(11):1396--1408, 1992.
\newblock \href {https://doi.org/10.1190/1.1443207}
  {\path{doi:10.1190/1.1443207}}.

\bibitem{Bechtold2011}
M.~Bechtold, S.~Haber-Pohlmeier, J.~Vanderborght, A.~Pohlmeier, T.~P.~A.
  Ferré, and H.~Vereecken.
\newblock Near-surface solute redistribution during evaporation.
\newblock {\em Geophysical Research Letters}, 38(17), 2011.
\newblock \href {https://doi.org/10.1029/2011GL048147}
  {\path{doi:10.1029/2011GL048147}}.

\bibitem{BringedalEigenvalue}
C.~Bringedal, G.~J.~M. Pieters, and C.~J. van Duijn.
\newblock {Eigenvalue problem solver for evaporation-driven density
  instabilities in saturated porous media}, 2022.
\newblock \href {https://doi.org/10.18419/darus-2577}
  {\path{doi:10.18419/darus-2577}}.

\bibitem{BringedalSatInstab}
C.~Bringedal, T.~Schollenberger, G.~J.~M. Pieters, C.~J. van Duijn, and
  R.~Helmig.
\newblock Evaporation-driven density instabilities in saturated porous media.
\newblock {\em Transport in Porous Media}, 143(2):297--341, 2022.
\newblock \href {https://doi.org/10.1007/s11242-022-01772-w}
  {\path{doi:10.1007/s11242-022-01772-w}}.

\bibitem{chaves2008plant}
M.~M. Chaves, J.~Flexas, and C.~Pinheiro.
\newblock {Photosynthesis under drought and salt stress: regulation mechanisms
  from whole plant to cell}.
\newblock {\em Annals of Botany}, 103(4):551--560, 07 2008.
\newblock \href {https://doi.org/10.1093/aob/mcn125}
  {\path{doi:10.1093/aob/mcn125}}.

\bibitem{chen1992evaporation}
X.~Y. Chen.
\newblock Evaporation from a salt encrusted sediment surface-field and
  laboratory studies.
\newblock {\em Soil Research}, 30(4):429--442, 1992.
\newblock \href {https://doi.org/10.1071/SR9920429}
  {\path{doi:10.1071/SR9920429}}.

\bibitem{salinityreview2016}
I.~N. Daliakopoulos, I.~K. Tsanis, A.~Koutroulis, N.~N. Kourgialas, A.~E.
  Varouchakis, G.~P. Karatzas, and C.~J. Ritsema.
\newblock The threat of soil salinity: A european scale review.
\newblock {\em Science of The Total Environment}, 573:727 -- 739, 2016.
\newblock \href {https://doi.org/10.1016/j.scitotenv.2016.08.177}
  {\path{doi:10.1016/j.scitotenv.2016.08.177}}.

\bibitem{dicarlo2013stability}
D.~A. DiCarlo.
\newblock Stability of gravity-driven multiphase flow in porous media: 40 years
  of advancements.
\newblock {\em Water Resources Research}, 49(8):4531--4544, 2013.
\newblock \href {https://doi.org/10.1002/wrcr.20359}
  {\path{doi:10.1002/wrcr.20359}}.

\bibitem{egorov2003stability}
A.~G. Egorov, R.~Z. Dautov, J.~L. Nieber, and A.~Y. Sheshukov.
\newblock Stability analysis of gravity-driven infiltrating flow.
\newblock {\em Water resources research}, 39(9), 2003.
\newblock \href {https://doi.org/10.1029/2002WR001886}
  {\path{doi:10.1029/2002WR001886}}.

\bibitem{Elenius2012}
M.~Elenius and K.~Johannsen.
\newblock On the time scales of nonlinear instability in miscible displacement
  porous media flow.
\newblock {\em Computational Geosciences}, 16, 09 2012.
\newblock \href {https://doi.org/10.1007/s10596-012-9294-2}
  {\path{doi:10.1007/s10596-012-9294-2}}.

\bibitem{GengInfluence}
X.~Geng and M.~C. Boufadel.
\newblock The influence of evaporation and rainfall on supratidal groundwater
  dynamics and salinity structure in a sandy beach.
\newblock {\em Water Resources Research}, 53(7):6218--6238, 2017.
\newblock \href {https://doi.org/10.1002/2016WR020344}
  {\path{doi:10.1002/2016WR020344}}.

\bibitem{gilman1996influence}
A.~Gilman and J.~Bear.
\newblock The influence of free convection on soil salinization in arid
  regions.
\newblock {\em Transport in Porous Media}, 23(3):275--301, 1996.
\newblock \href {https://doi.org/10.1007/BF00167100}
  {\path{doi:10.1007/BF00167100}}.

\bibitem{hattori2015robin}
T.~Hattori, J.~C. Patterson, and C.~Lei.
\newblock On the stability of transient penetrative convection driven by
  internal heating coupled with a thermal boundary condition.
\newblock {\em International Communications in Heat and Mass Transfer},
  64:29--33, 2015.
\newblock \href {https://doi.org/10.1016/j.icheatmasstransfer.2015.02.009}
  {\path{doi:10.1016/j.icheatmasstransfer.2015.02.009}}.

\bibitem{Helmig2011}
R.~Helmig.
\newblock Multiphase flow and transport processes in the subsurface: A
  contribution to the modeling of hydrosystems.
\newblock 2011.
\newblock URL: \url{https://api.semanticscholar.org/CorpusID:118931613}.

\bibitem{jambhekar}
V.~A. Jambhekar, R.~Helmig, N.~Schröder, and N.~Shokri.
\newblock Free-flow–porous-media coupling for evaporation-driven transport
  and precipitation of salt in soil.
\newblock {\em Transport in Porous Media}, 110(2):251–280, 11 2015.
\newblock \href {https://doi.org/10.1007/s11242-015-0516-7}
  {\path{doi:10.1007/s11242-015-0516-7}}.

\bibitem{KlokerCode}
L.~Kloker and C.~Bringedal.
\newblock {Code for: Solution approaches for evaporation-driven density
  instabilities in a slab of saturated porous media}, 2022.
\newblock \href {https://doi.org/10.18419/darus-3057}
  {\path{doi:10.18419/darus-3057}}.

\bibitem{KlokerApproaches}
L.~H. Kloker and C.~Bringedal.
\newblock Solution approaches for evaporation-driven density instabilities in a
  slab of saturated porous media.
\newblock {\em Physics of Fluids}, 34(9):096606, 2022.
\newblock \href {https://doi.org/10.1063/5.0110129}
  {\path{doi:10.1063/5.0110129}}.

\bibitem{Koch2021}
T.~Koch, D.~Gläser, K.~Weishaupt, S.~Ackermann, M.~Beck, B.~Becker,
  S.~Burbulla, H.~Class, E.~Coltman, S.~Emmert, T.~Fetzer, C.~Grüninger,
  K.~Heck, J.~Hommel, T.~Kurz, M.~Lipp, F.~Mohammadi, S.~Scherrer,
  M.~Schneider, G.~Seitz, L.~Stadler, M.~Utz, F.~Weinhardt, and B.~Flemisch.
\newblock Dumux 3 – an open-source simulator for solving flow and transport
  problems in porous media with a focus on model coupling.
\newblock {\em Computers \& Mathematics with Applications}, 81:423--443, 2021.
\newblock Development and Application of Open-source Software for Problems with
  Numerical PDEs.
\newblock \href {https://doi.org/10.1016/j.camwa.2020.02.012}
  {\path{doi:10.1016/j.camwa.2020.02.012}}.

\bibitem{lasser2021stability}
J.~Lasser, M.~Ernst, and L.~Goehring.
\newblock Stability and dynamics of convection in dry salt lakes.
\newblock {\em Journal of Fluid Mechanics}, 917:A14, 2021.
\newblock \href {https://doi.org/10.1017/jfm.2021.225}
  {\path{doi:10.1017/jfm.2021.225}}.

\bibitem{emna2017}
E.~Mejri, R.~Bouhlila, and R.~Helmig.
\newblock Heterogeneity effects on evaporation-induced halite and gypsum
  co-precipitation in porous media.
\newblock {\em Transport in Porous Media}, 118:39--64, 2017.
\newblock \href {https://doi.org/10.1007/s11242-017-0846-8}
  {\path{doi:10.1007/s11242-017-0846-8}}.

\bibitem{Millington1961}
R.~J. Millington and J.~P. Quirk.
\newblock {Permeability of porous solids}.
\newblock {\em Trans. Faraday Soc.}, 57:1200--1207, 1961.
\newblock \href {https://doi.org/10.1039/TF9615701200}
  {\path{doi:10.1039/TF9615701200}}.

\bibitem{nieldbejan2017}
D.~A. Nield and A.~Bejan.
\newblock {\em Convection in porous media}.
\newblock Springer, 5 edition, 2017.

\bibitem{piortrowski2020crust}
J.~Piotrowski, J.~A. Huisman, U.~Nachshon, A.~Pohlmeier, and H.~Vereecken.
\newblock Gas permeability of salt crusts formed by evaporation from porous
  media.
\newblock {\em Geosciences}, 10(11), 2020.
\newblock \href {https://doi.org/10.3390/geosciences10110423}
  {\path{doi:10.3390/geosciences10110423}}.

\bibitem{pitman2020salinity}
M.~G. Pitman and A.~Läuchli.
\newblock Global impact of salinity and agricultural ecosystems.
\newblock In A.~Läuchli and U.~Lüttge, editors, {\em Salinity:
  Environment-Plants-Molecules}, pages 3--20. Springer, 2002.

\bibitem{Riaz2006}
A.~Riaz, M.~Hesse, H.~A. Tchelepi, and F.~M. Orr.
\newblock Onset of convection in a gravitationally unstable diffusive boundary
  layer in porous media.
\newblock {\em Journal of Fluid Mechanics}, 548:87–111, 2006.
\newblock \href {https://doi.org/10.1017/S0022112005007494}
  {\path{doi:10.1017/S0022112005007494}}.

\bibitem{Darus_2020}
M.~Schneider, B.~Flemisch, S.~Frey, S.~Hermann, D.~Iglezakis, M.~Ruf,
  B.~Schembera, A.~Seeland, and H.~Steeb.
\newblock Datenmanagement im sfb 1313.
\newblock {\em Bausteine Forschungsdatenmanagement}, (1):28–38, Apr. 2020.
\newblock \href {https://doi.org/10.17192/bfdm.2020.1.8085}
  {\path{doi:10.17192/bfdm.2020.1.8085}}.

\bibitem{singh2015}
K.~Singh.
\newblock Microbial and enzyme activities of saline and sodic soils.
\newblock {\em Land Degradation \& Development}, 27(3):706--718, 2016.
\newblock \href {https://doi.org/10.1002/ldr.2385}
  {\path{doi:10.1002/ldr.2385}}.

\bibitem{Slim_2014}
A.~C. Slim.
\newblock Solutal-convection regimes in a two-dimensional porous medium.
\newblock {\em Journal of Fluid Mechanics}, 741:461–491, 2014.
\newblock \href {https://doi.org/10.1017/jfm.2013.673}
  {\path{doi:10.1017/jfm.2013.673}}.

\bibitem{ursino2000linear}
N.~Ursino.
\newblock Linear stability analysis of infiltration, analytical and numerical
  solution.
\newblock {\em Transport in porous media}, 38(3):261--271, 2000.
\newblock \href {https://doi.org/10.1023/A:1006688232755}
  {\path{doi:10.1023/A:1006688232755}}.

\bibitem{van2004steady}
C.~J. van Duijn, G.~J.~M. Pieters, and P.~A.~C. Raats.
\newblock Steady flows in unsaturated soils are stable.
\newblock {\em Transport in Porous Media}, 57(2):215--244, 2004.
\newblock \href {https://doi.org/10.1023/B:TIPM.0000038250.72364.20}
  {\path{doi:10.1023/B:TIPM.0000038250.72364.20}}.

\bibitem{vanduijn2019stability}
C.~J. van Duijn, G.~J.~M. Pieters, and P.~A.~C. Raats.
\newblock On the stability of density stratified flow below a ponded surface.
\newblock {\em Transport in Porous Media}, 127(3):507--548, 2019.
\newblock \href {https://doi.org/10.1007/s11242-018-1209-9}
  {\path{doi:10.1007/s11242-018-1209-9}}.

\bibitem{van2001stability}
C.~J. van Duijn, R.~A. Wooding, G.~J.~M. Pieters, and A.~van~der Ploeg.
\newblock {\em Stability criteria for the boundary layer formed by throughflow
  at a horizontal surface of a porous medium}.
\newblock American Geophysical Union (AGU), 2002.

\bibitem{vanGenuchten1980}
M.~T. van Genuchten.
\newblock A closed-form equation for predicting the hydraulic conductivity of
  unsaturated soils.
\newblock {\em Soil Science Society of America Journal}, 44(5):892--898, 1980.
\newblock \href {https://doi.org/10.2136/sssaj1980.03615995004400050002x}
  {\path{doi:10.2136/sssaj1980.03615995004400050002x}}.

\bibitem{WoodingSalt1}
R.~A. Wooding, S.~W. Tyler, and I.~White.
\newblock Convection in groundwater below an evaporating salt lake: 1. onset of
  instability.
\newblock {\em Water Resources Research}, 33(6):1199--1217, 1997.
\newblock \href {https://doi.org/10.1029/96WR03533}
  {\path{doi:10.1029/96WR03533}}.

\bibitem{WoodingSalt2}
R.~A. Wooding, S.~W. Tyler, I.~White, and P.~A. Anderson.
\newblock Convection in groundwater below an evaporating salt lake: 2.
  evolution of fingers or plumes.
\newblock {\em Water Resources Research}, 33(6):1219--1228, 1997.
\newblock \href {https://doi.org/10.1029/96WR03534}
  {\path{doi:10.1029/96WR03534}}.

\end{thebibliography}

\end{document}